\documentclass[useAMS,usenatbib]{mn2e}

\usepackage{graphicx,amssymb,hyperref}
\usepackage[fleqn]{amsmath}  
\usepackage{multirow}
\usepackage{widetext}
\usepackage{pifont}

\usepackage{graphicx,amssymb,multirow}
\usepackage[fleqn]{amsmath}  

\def\simlt{\lower.5ex\hbox{$\; \buildrel < \over \sim \;$}}
\def\simgt{\lower.5ex\hbox{$\; \buildrel > \over \sim \;$}}
\def\etal{{\it et al.}}

\def\eg{{\it e.g.}}

\def\acswfc{{\sl ACS/WFC}}
\usepackage{graphicx} 
\newcommand{\be}{\begin{equation}}
\newcommand{\ee}{\end{equation}}
\newcommand{\ba}{\begin{eqnarray}}
\newcommand{\ea}{\end{eqnarray}}

\usepackage{graphicx} 

\title[Improved CTI correction for HST]{
An improved model of Charge Transfer Inefficiency and correction algorithm for the Hubble Space Telescope}
\author[R.\ Massey \etal]{Richard Massey$^{1}$\thanks{e-mail: {\tt r.j.massey@durham.ac.uk}}, 
Tim Schrabback$^{2}$, Oliver Cordes$^{2}$, Ole Marggraf$^{2}$, Holger \newauthor 
Israel$^{1}$, Lance Miller$^{3}$, David Hall$^{4}$, Mark Cropper$^{5}$, Thibaut Prod'homme$^{6}$ \newauthor 
and Sami-Matias Niemi$^{5}$ \\
$^{1}$ Institute for Computational Cosmology, Durham University, South Road, Durham DH1 3LE, UK\\ 
$^{2}$ Argelander-Institut f\"ur Astronomie, Auf dem H\"ugel 71, D-53121 Bonn, Germany\\ 
$^{3}$ Department of Physics, University of Oxford, The Denys Wilkinson Building, Keble Road, Oxford, OX1 3RH, UK\\ 
$^{4}$ Centre for Electronic Imaging, Department of Physical Sciences, The Open University, Walton Hall, Milton Keynes MK7 6AA, UK\\ 
$^{5}$ Mullard Space Science Laboratory, University College London, Holmbury St Mary, Dorking, Surrey RH5 6NT, UK\\ 
$^{6}$ European Space Agency, ESTEC, Postbus 299, NL-2200 AG Noordwijk, the Netherlands\\
\vspace{-5mm}}
\begin{document}
\date{Accepted 2014 January 2. Received 2013 December 20; in original form 2013 October 14.}

\pagerange{\pageref{firstpage}--\pageref{lastpage}} \pubyear{2014}

\maketitle

\label{firstpage}

\begin{abstract}

Charge-Coupled Device (CCD) detectors, widely used to obtain digital imaging, can be damaged by high energy radiation.
Degraded images appear blurred, because of an effect known as Charge Transfer Inefficiency (CTI), which trails bright objects as the image is read out.
It is often possible to correct most of the trailing during post-processing, by moving flux back to where it belongs.
We compare several popular algorithms for this: quantifying the effect of their physical assumptions and tradeoffs between speed and accuracy.
We combine their best elements to construct a more accurate model of damaged CCDs in the {\it Hubble Space Telescope}'s {\it Advanced Camera for Surveys}/{\it Wide Field Channel}, and update it using data up to early 2013.
Our algorithm now corrects 98\% of CTI trailing in science exposures, a substantial improvement over previous work.
Further progress will be fundamentally limited by the presence of read noise.
Read noise is added after charge transfer so does not get trailed -- but it is incorrectly {\em untrailed} during post-processing.

\end{abstract}


\section{Introduction}

The harsh radiation environment above the Earth's atmosphere, in particle accelerators, or in medical contexts, gradually degrades Charge-Coupled Device (CCD) imaging detectors.
CCD detectors work by converting photons to electrons inside a silicon lattice, then collecting the electrons in electrostatic potential wells that form each pixel. 
At the end of an exposure, the photoelectrons are transferred through a chain of pixels to the edge of the device, where they are amplified for external counting.
However, radiation damage to the silicon lattice creates charge traps that capture electrons for a short time. 
When electrons are delayed during their transfer to the amplifier, the effect is known as `Charge Transfer Inefficiency' (CTI). 
Those electrons emerge several pixels later, as spurious trails behind every source \citep{holland90,janesick01}.
Different `species' of traps (different configurations of the damaged silicon lattice) capture electrons for different lengths of time, resulting in a variety or complex superposition of trail profiles.

CTI trailing is particularly troublesome because the amount of flux trailed is a nonlinear function of the flux, size and shape of a source, and the previous illumination history. 
The effect on images is therefore {\it not} a simple shape transformation such as might be described by a convolution. 
To first order, this can be roughly understood by considering that there are only a finite number of charge traps, so the {\it fraction} of electrons trailed from faint sources will be greater than that from bright sources. 
The process of adding CTI trails can be reproduced in software that models the flow of electrons past the charge traps. 

\begin{figure}
\begin{center}
\includegraphics[width=70mm]{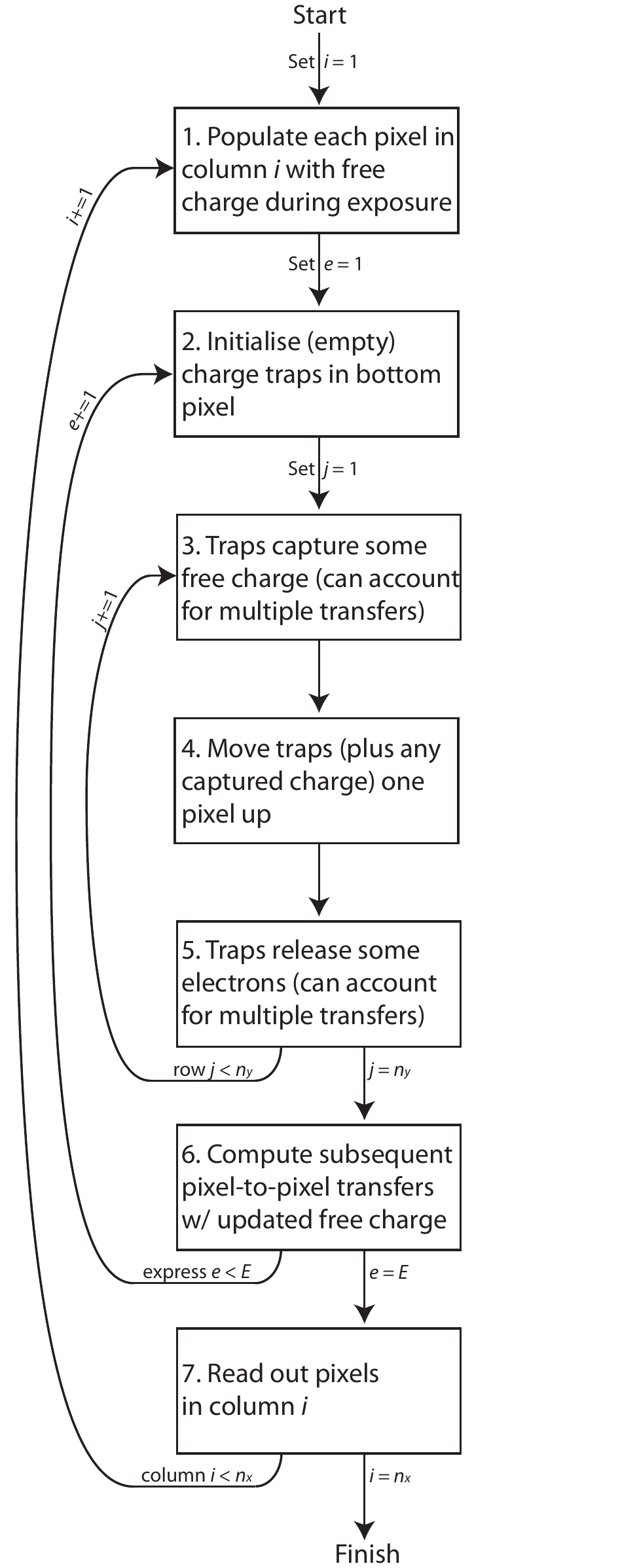}
\end{center}
\caption{A generic forward algorithm to add CTI trailing to an image, mimicking the effects of radiation-damaged hardware.
The number of pixel columns and rows is $n_x\times n_y$, while the parameter $E$ governs a tradeoff between speed and accuracy that is discussed in section~\ref{sec:balance}.
All CTI correction algorithms currently in operation on astronomical telescopes follow this same framework.
}
\label{fig:flowchart}
\end{figure}

Since trailing is (almost) the last process to happen during data acquisition, it can be corrected during the first stages of data analysis.
\citet{bristow03im} pioneered an iterative method to reverse the effect of CCD readout in images from the {\it Hubble Space Telescope} ({\sl HST}).
With this technique, CTI trailing can be removed by repeatedly running the software to {\em add} CTI trailing. 
\cite{m10a} used this technique to correct 90\% of the trailing in {\sl HST} {\it Advanced Camera for Surveys}/{\it Wide Field Channel} (\acswfc) data from 2006, and the approach was incorporated into the STScI CALACS software by \cite{a10}.
As the radiation damage accumulated, the trailing increased and became easier to measure.
\cite{m10b} updated the model and achieved a 95\% correction in data from 2010.

Although most of the trailing can now be removed, the residual will still limit high precision science with {\sl HST}.
There are many examples where measurements of photometry, astrometry and morphology are required to better than 5\% \citep[e.g.][]{snephot,capak07,propmot,ghez08,rago12,great10gal} -- and {\sl HST}'s detectors continue to degrade.
The same effect will also limit the European Space Agency (ESA)'s future missions {\it Gaia} \citep{gaia} and {\it Euclid} \citep{euclid}.
For example, {\it Euclid} will require 99\% correction \citep{m13,cropper13}.
In this paper, we attempt to build a new model for {\sl HST} \acswfc\ that meets this level of accuracy.

This paper is organised as follows.
In section~\ref{sec:algorithm}, we study the details of three popular algorithms to correct CTI, comparing their physical assumptions and their approximations made for speed.
In section~\ref{sec:newmodel},we measure relevant properties of the {\sl HST} \acswfc\ detectors using real data, and quantitatively compare the algorithmic options using simulated data (where the truth is known). 
In section~\ref{sec:choices}, we identify suitable algorithmic choices to achieve high accuracy in a minimum of time, and test
our improved correction on real data in section~\ref{sec:results}.
In section~\ref{sec:conclusions}, we summarise lessons learned, and discuss possible avenues for future work.

\section{Detailed code comparison} \label{sec:algorithm}

CTI correction splits naturally into two tasks.
It is first necessary to be able to add trails by mimicking the hardware process that happens during readout (section~\ref{sec:algorithm_add}).
Attempting to then undo the trailing is a separate process (section~\ref{sec:algorithm_subtract}).
We shall compare approaches taken in the literature, attempting to understand their unique advantages and disadvantages -- and aiming to combine their best practices.

\subsection{Adding CTI trailing} \label{sec:algorithm_add}

In recent years, a series of ever-more sophisticated algorithms have been developed to model charge trapping in CCD readout.
Three codes are now in widespread use, each adopting slightly different model assumptions.
The \citet{m10a} and \citet{a10} codes were developed for {\sl HST} and generally aim for accuracy at the expense of computation speed.
They share a common heritage, but considerable development has been put in since they forked.
The \citet{short10,short13} code was developed independently for {\it Gaia}, and is built for speed.

All three algorithms share the same basic framework for adding CTI trails, which is illustrated for parallel CTI\footnote{To add serial CTI, \citet{rhodes10}\ rotated the image by 90 degrees and rerunning the same code with a different trapping model. 
This isn't perfectly accurate, because parallel CTI codes generally do not model the illumination history from the (long ago) previous exposure. 
The history relevant for serial CTI is the trap occupancy from readout of the row below, which passed through the serial register {\it immediately} before. 
Serial CTI in {\sl HST}/{\sl ACS} is very small, and we do not consider it in this paper.} in figure~\ref{fig:flowchart}.
The different choices made for various parameters are listed in table~\ref{tab:metamodel}, but the independent codes share some remarkable similarities.
For example, while in hardware, electrons are transferred down through traps to a readout node, in software all the algorithms scan traps up through the electrons.
This change of `inertial reference frame' saves computational overheads by keeping the main image array static.

The core components of each method are two functions: one controlling which traps capture an electron (box 3), and one controlling when the electrons are released (box 5). 
Fundamental differences also arise in the accounting practices adopted to monitor which traps are full at any time, and in approximations used to speed up the inner loop (boxes 3--5).
We shall discuss each of these steps in more detail.

\begin{table*}
 \centering
  \caption{Summary of the parameter choices made for various forward algorithms that add CTI trailing to an image. 
  The core framework shared by all algorithms is to: capture electrons in traps, move the electrons (or traps), then release some electrons.
  Different approximations have been made in these three steps, and in accounting practices to monitor which traps are occupied.
  Note that, where numbers are $>$1 or $<$$n_y$ (the number of rows of pixels in a CCD), they are user-configurable and have been varied to fit the CCD in question.
  \label{tab:metamodel}}
  \begin{tabular}{lccr@{$\times$}lcccc}
                & Trap & Trap & \multicolumn{2}{c}{Trap monitoring} & Transfer many at & Transfer thro' & Trap & Serial \\
Algorithm & initialisation & capture & \multicolumn{2}{c}{$n_\mathrm{species}$$\times$$n_\mathrm{levels}$} & once ($P$=$n_y/E$) & gaps ($n_\mathrm{phases}$) & release & transfer \\
  \hline
\cite{bristow03im} & all empty & $\quad$ $\beta$, $d$ & 3 & discrete$^a$ & 1 & 3 & $\tau$ & $\checkmark$ \\
\cite{m10a} & all empty & $\quad$ $\beta$, $d$ & $\quad$ 3 & discrete$^b$ & 1 & -- & $\tau$ & -- \\
\cite{rhodes10} & all empty & $\quad$ $\beta$, $d$ & 5 & 10,000 & 1 & 3 & $\tau$ & $\checkmark$ \\
\cite{a10} & all empty & lookup & 1 & $\rho_\mathrm{t}V_\mathrm{pix}n_y$ & $n_y$ & -- & lookup & -- \\
\cite{m10b} & all empty & $\beta$ & 3 & 10,000 & $n_y$/3 & -- & $\tau$ & -- \\
\cite{hall10} & all empty & full sim & 5 & discrete$^a$ & 1 & full sim & $\tau$ & -- \\
\cite{short13} & nearby objs & $\alpha$, $\beta$ $\quad$ & 4 & 1 & $n_y$ & -- & $\tau$ & -- \\
 \hline
 This work & all empty & $\beta$ & 3 & $\rho_\mathrm{t}V_\mathrm{pix}n_y$ & $n_y$/5 & -- & $\tau$ & -- \\
\hline
\multicolumn{9}{r}{\scriptsize{$^a$A population of traps is scattered throughout the model CCD.~~
$^b$The population of traps is oversampled, each trap holding a fraction of an electron.}} 
\end{tabular}
\end{table*}

\subsubsection{Initialising charge traps (box 2)}

To model parallel CTI, the charge traps are generally created empty.
After a long, static integration on sky, most traps will have had ample opportunity to release electrons.
In a few regions of the CCD, a steady state may have been reached in which some traps are full.
However, the first operation during a readout algorithm is for traps to capture electrons.
This sets occupancy to the desired levels before readout continues, rendering as immaterial their occupancy during integration (if capture is instantaneous, which is assumed for volume-driven models -- see below).

One notable exception to this is the {\it Gaia} observations modelled by \cite{short13}.
These observations are performed in `Time-Delay Integration' mode, in which the shutter is left open and electrons are continuously read out at the same rate as the telescope is slewed.
When initialising traps, it is therefore necessary to consider nearby sources that may have recently transited the focal plane and pre-filled traps \citep{prod12}
(this is exacerbated if capture is not instantaneous, as assumed in density-driven models -- see below).

\subsubsection{Capture of electrons into charge traps (box 3)}

To determine which traps can capture an electron, we start by considering the location of the traps and the electrons.

Charge traps of all species are scattered throughout a CCD's 3D silicon lattice.
In-orbit measurements show that each species of traps in {\sl ACS}/{\sl WFC} have uniform density $\rho_\mathrm{t}$ \citep[][section~2.5]{m10a}, and are not significantly clustered (\citealt{ogaz13} show column-to-column shot noise in $\rho_\mathrm{t}$ that appears Poissonian).

\begin{figure}
\begin{center}
\includegraphics[width=83mm]{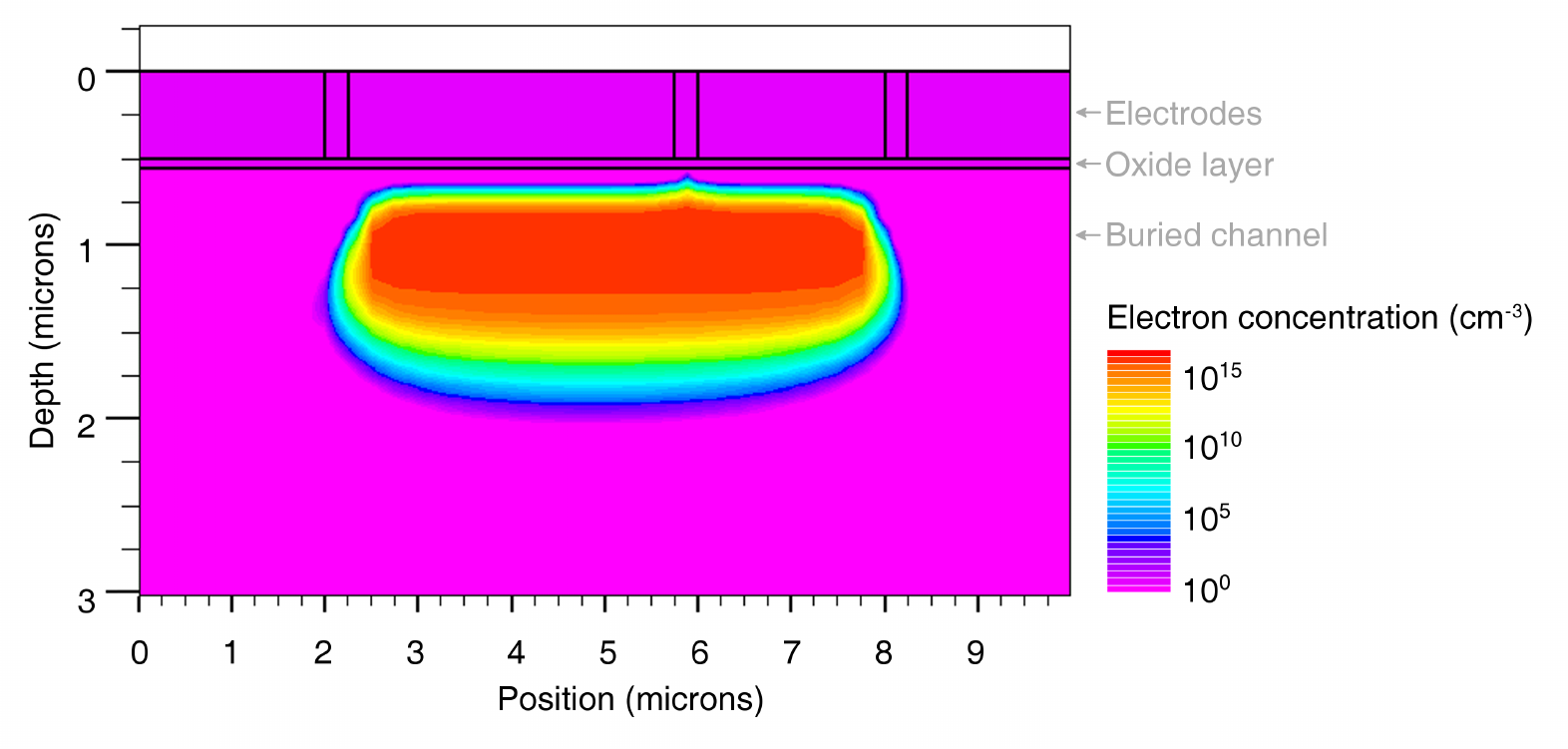} \\
\includegraphics[width=41mm]{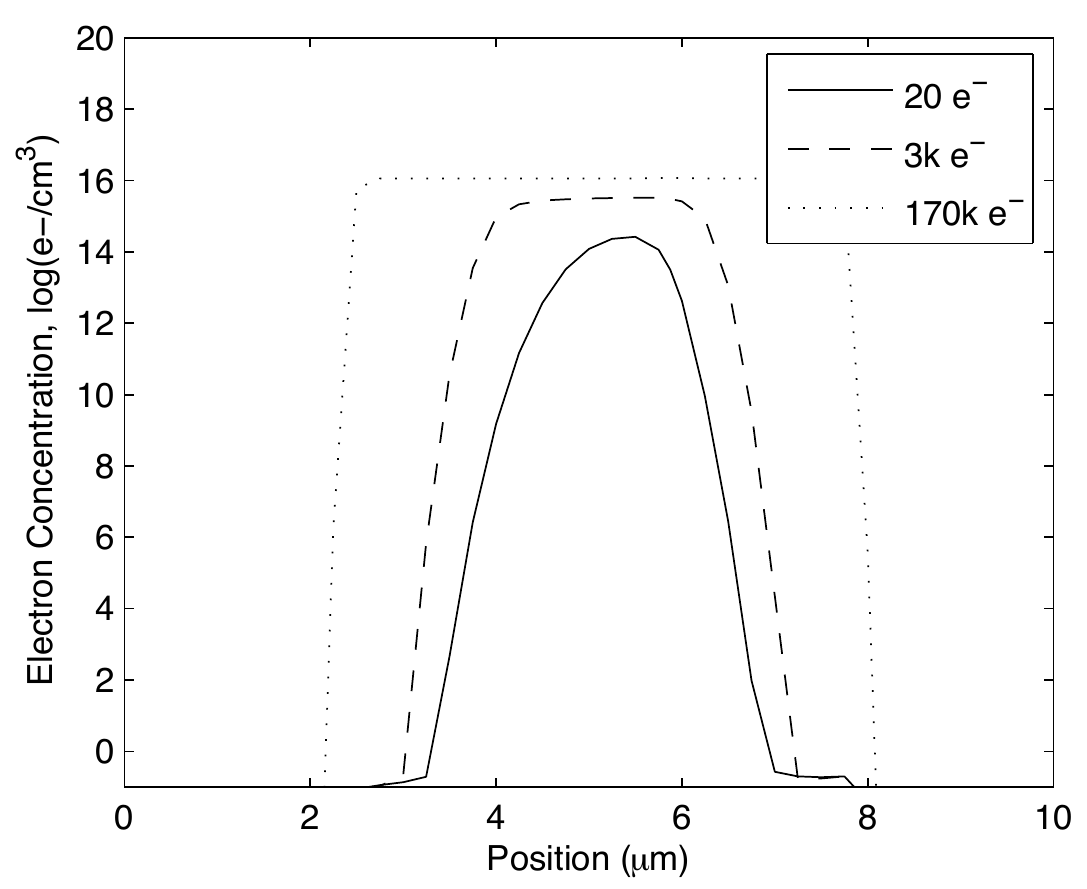}
\includegraphics[width=41mm]{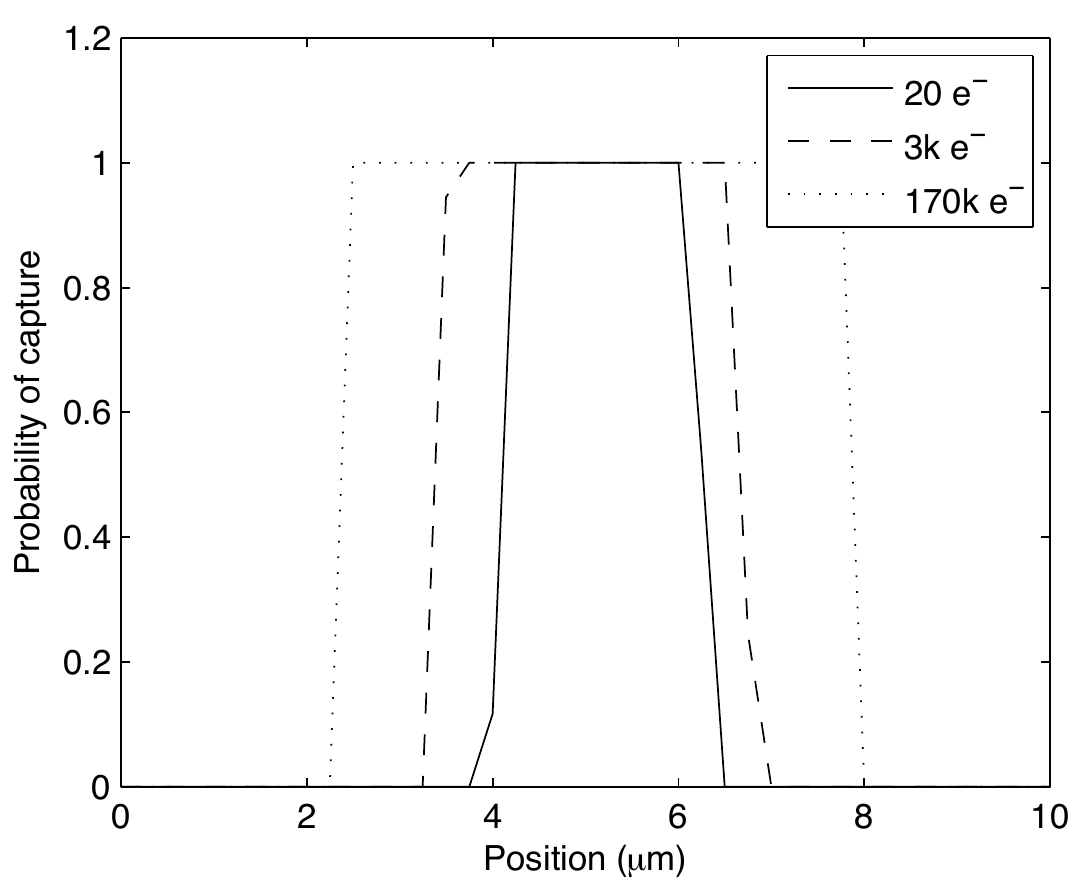} \\
\end{center}
\caption{
{\it Top}: 
the density of electrons within one pixel of a generic CCD.
This is a two-dimensional slice through the full three-dimensional ATLAS model.
The charge cloud in this particular example contains $172,000$\ electrons, and the detector model includes a 4-phase, uneven electrode structure for readout in the left-right direction on the page.
{\it Bottom-left}:
a one-dimensional slice through the electron density.
{\it Bottom-right}:
the probability that an electron will be captured by a trap at that location, computed by combining the electron density with the capture cross-section for a specific trap species (here Si-E) and a set dwell-time (here $1$\,ms), through the use of Shockley-Read-Hall theory.}
\label{fig:silvaco}
\end{figure}

The distribution of electrons $\rho_e(\mathbf{x})$ is more complex.
For storage and transportation, electrons are confined to a `buried channel' (BC) region of the silicon lattice by an electric field, which can be shaped by doping the silicon with other elements during manufacture.
Information about the precise doping structures is industrially sensitive, but \citet{seabroke} and \citet{clarke12a, clarke12b} use ATLAS semiconductor device simulation software \citep{silvaco} to model the geometry and volume of a cloud of $n_e$ electrons in generic devices (top panel in figure~\ref{fig:silvaco}).
The volume $V_e(n_e)$ within any electron density threshold typically grows as $V_e\propto n_e^\beta + c$ \citep{hall10}, and converges to the geometric volume of (the BC in) a pixel $V_\mathrm{pix}$ as $n_e\rightarrow w$, the full well depth (bottom-left panel in figure~\ref{fig:silvaco}).
The constant $c$ prevents the volume of a cloud of electrons tending to unphysically small values for low $n_e$.

During each clock cycle (pixel-to-pixel transfer) of CCD readout, the electrons are held stationary for a time $t_\mathrm{dwell}$.
During this time, the probability $p_\mathrm{capture}(\mathbf{x})$ that a trap at position $\mathbf{x}$ captures an electron depends upon $\rho_e(\mathbf{x})$ and the capture cross-section of the trap $\sigma_\mathrm{t}$ \citep[][]{srhsr,srhh}.
The capture cross-section varies between trap species and also depends on operating conditions like the temperature $T$.
Integrating over all the traps (all the volume) within a pixel, the total number of electrons that will be captured into each species of charge trap is
\be
n_\mathrm{c}= \hspace{-1mm} 
\int_0^{V_\mathrm{pix}}  \hspace{-2mm} \mathrm{d}V ~ \rho_\mathrm{t}^\mathrm{empty}(\mathbf{x}) ~ p_\mathrm{capture}\Big(\rho_e(\mathbf{x};n_e),\sigma_\mathrm{t}(t_\mathrm{dwell},T)\Big) 
\label{eqn:fullnc}
\ee
where only the initially empty traps (or fractions of traps) are considered in the trap density $\rho_\mathrm{t}^\mathrm{empty}(\mathbf{x})$.

\vspace{4mm}

ATLAS models suggest that the probability of capture during a dwell time $\sim$$1$\,ms is typically close to a step function in the electron density.
That is, inside some density threshold $p_\mathrm{capture}$$\approx$1 and all the traps capture an electron, but outside that threshold, $p_\mathrm{capture}$$\approx$0 (bottom-right panel in figure~\ref{fig:silvaco}).
Empirical tests with in-orbit data confirm that this step function is a suitable approximation for {\sl ACS}/{\sl WFC} \citep[][section~3.2]{a10}.
This simplifies equation~\eqref{eqn:fullnc}: $p_\mathrm{capture}(\rho_e,\sigma_\mathrm{t})$ becomes a delta function inside $V_e$, which is realised as a volume when the integration is done.
Since $\rho_\mathrm{t}$ is constant, it can also be moved outside the integral.
This leads to a `volume-driven' model, in which the number of electrons captured during each dwell time,
\be
n_\mathrm{c}(n_e) = \rho^\mathrm{empty}_\mathrm{t}~V_e(n_e),
\ee
depends solely on the effective volume $V_e(n_e)$ of the packet.
A packet of only few electrons is confined to a small volume of silicon, and thus `sees' to only a fraction of the charge traps.
As the number of electrons $n_e$ grows or shrinks, so does the number of exposed traps.

If a cloud of electrons must traverse $y$ pixels to reach the readout register, the volume of silicon (and the number of traps) to which they are exposed increases by this factor.
Incorporating that factor of $y$ (but we shall return a more sophisticated treatment of this issue in section~\ref{sec:balance}), the \citet{m10b} algorithm empirically fitted a form 
\be
n_\mathrm{c}(n_e)=\mathrm{min}\bigg\{\rho^\mathrm{empty}_\mathrm{t}y ~ V_\mathrm{pix}\bigg( \frac{n_e}{w} \bigg)^\beta ,~n_ey\bigg\},
\label{eqn:m10vol}
\ee
where the min\{\} explicitly ensures the number of electrons stays positive, and $\beta\approx0.5$.
Like the \citet{hall10} fitting function, the code also allows $V_e(n_e)$ to include a constant.
In early analysis, \cite{m10a} (their equation~4) found a slightly negative best-fit constant, and interpreted it as tentative evidence for a supplementary buried channel (SBC).
It remains unknown whether the \acswfc\ CCDs have an SBC (J.\ Anderson 2013, pers.\ comm.): one was designed in the silicon doping structure, but it may never have properly formed, or evaporated away after manufacture. 
The constant was subsequently set to zero in \citet{m10b}, because the data analysed contained a relatively bright sky background so was insensitive to it.

The \citet{a10} algorithm is also volume-driven, but uses an empirical look-up table for $n_\mathrm{c}(n_e)$ instead of an analytic function.
They analysed data with lower sky backgrounds, and found a relative {\em decrease} in the number of traps exposed to small (DN$_2$=10) electron clouds \citep[][figure~6]{a10} that supports the existence of a SBC.

\vspace{4mm}

Not assuming that $p_\mathrm{capture}$ is a step function led \citet{short13} instead to a partially `density-driven' algorithm.
This is particularly important because the {\it Gaia} CCDs will often need to transfer very small numbers of electrons between pixels (because in TDI observations, the exposure time for the first pixel-to-pixel transfers is almost nil).
In a density-driven model, a packet of a few electrons is assumed to occupy a relatively larger\footnote{The volume of electrons $V_e\propto (n_e/w)^\beta$ increases as $\beta$ decreases because $n_e/w<1$.} volume ($\beta$$\sim$$0.3$) and `see' more traps.
However, with a low electron density inside that larger volume, only a fraction of the traps capture an electron.
The fraction is governed by a parameter $\alpha$, which varies between trap species and depends on $\sigma_{t}(t_\mathrm{dwell},T)$.
\citet{prod12} and \citet{short13} used this approach to reproduce lab test data in a much wider range of operating conditions and lower signals than is typically encountered by \acswfc.

The number of captures gains an extra term
\be \label{eqn:densdriven1}
n_\mathrm{c}(n_e)=\frac{2\rho_\mathrm{t} y V_\mathrm{pix} (n_e/w)^\beta}{2\rho_\mathrm{t} y V_\mathrm{pix} (n_e^{\beta-1}/w^\beta)+1}\bigg( 1-\mathrm{exp} \Big\{-\alpha~n_e^{1-\beta} \Big\} \bigg)
\ee
\citep[c.f.][equation~22, assuming that the traps do indeed start empty]{short13}.
Note that the factor 2 (and the counterintuitive appearance of $y$ in the denominator) arise from a peculiar approximation of $\mathrm{min}\{A,B\}\approx AB/(A+B)$ in \citet{short13} equation~5.
There is no reason for this, and is more physically intuitive rewritten as
\be \label{eqn:densdriven2}
n_\mathrm{c}(n_e)=\mathrm{min}\bigg\{\rho_\mathrm{t} y V_\mathrm{pix} \bigg(\frac{n_e}{w}\bigg)^\beta,~n_ey\bigg\}\bigg( 1-\mathrm{exp} \Big\{-\alpha~n_e^{1-\beta} \Big\} \bigg)
\ee
where we see the first term identically recovers equation~\eqref{eqn:m10vol}.

\subsubsection{Monitoring trap occupancy}\label{sec:nlevels}

Monitoring which traps are full at any time during the many pixel-to-pixel transfers requires an accounting scheme.
Early codes \citep[e.g][]{bristow03im} modelled populations of discrete traps, each able to capture a single electron, and kept a ledger of their occupancy.
This would be ideal if the model traps could be placed at the true trap positions (which could in principle be measured by in-orbit trap-pumping, \citealt{snap_pocket}).
However, the position of traps in {\sl HST} CCDs are unknown.
To implement this approach, model trap positions would therefore have to be scattered at random in 3D.
Every random realisation of trap positions (in both hardware and software) adds shot noise to the trails.
The random locations of traps in the hardware is unavoidable but, to remove the software half of this noise, \citet{rhodes10} instead modelled a fixed grid of $n_\mathrm{levels}$ fractional traps.
Each fractional trap is able to hold up to $\rho_\mathrm{t} V_{\mathrm pix}/n_\mathrm{levels}$ fractions of an electron, and samples a small volume within a pixel. Because $V_e$ grows monotonically with $n_e$ (see figure~\ref{fig:silvaco}), these small volumes can represent successive regions of a pixel into which a cloud of electrons will next grow, and the 3D integral in equation~\eqref{eqn:fullnc} can be converted to a 1D integral over the ordered set of traps.

This approach has been followed ever since, but the adopted density of fractional traps varies considerably between modern algorithms.
At the most dense extreme, the \citet{m10b} algorithm monitors an array of $3\times$10,000 traps in every pixel.
The $n_{\mathrm{species}}$=3 trap species observed in {\sl HST} each delay electrons by a different characteristic time.
After a coarse convergence test, $n_\mathrm{levels}$=10,000 equal-sized traps in every pixel were found necessary to sample the volume within each pixel.
Such an enormous array is slow to manipulate (execution time scales roughly as the total number of traps to the power 0.55), so the other codes each use an approximation to shrink one of its dimensions.

The \citet{short13} algorithm monitors $n_{\mathrm{species}}$=$4$ species of traps that contribute to parallel CTI \citep{prod10} -- but the entire volume of the CCD pixel is bundled together.
Only a single number is monitored for each trap species in each pixel: the total fraction of currently occupied traps
This less detailed accounting saves a factor of 10,000 traps and provides a significant speed boost, but may not be sufficiently accurate for our purposes.

The \citet{a10} algorithm monitors traps at many $n_\mathrm{levels}$ within the CCD -- but the different species are combined into one `composite' type.
The composite traps represent a mixture of the three physical species of traps, so they have complex behaviour (not exponential release), and their behaviour may also vary at different heights within the CCD.

\citet{a10} make two especially useful points about the trap accounting.
First, although intermediate calculations deal with fractions of electrons, the number of electrons in each pixel after readout (and the file formats that store these data) must be integers. 
Achieving greater precision is therefore unnecessary.
Staying within a precision of 1 electron after $n_y$ transfers requires only $n_\mathrm{levels}\ge\rho_\mathrm{t}V_\mathrm{pix}n_y$ traps, each of which interact with one marginal electron. 
Second, the traps need not be all the same size.
Where marginal electrons occupy a different volume, it could be most efficient to use large/small traps holding many/few electrons.
Indeed, in accounting terms, the traps could be spaced linearly in terms of volume, marginal free electrons, or marginal exposed traps.

\subsubsection{Pixel-to-pixel parallel transfers (box 4)}

During every cycle of pixel-to-pixel parallel transfers, electrons still in the image array are held stationary for a dwell time $t_\mathrm{dwell}$, while electrons in the serial register are transferred to the readout electronics.
At the end of the cycle, electrons in the image array are then moved rapidly to an adjacent pixel.
They pass fleetingly through the volume of silicon between pixels (i.e.\ at position $<$$2\mu m$ or $>$$8\mu m$ in the top panel of figure~\ref{fig:silvaco}).
The time spent in this volume is orders of magnitude shorter than $t_\mathrm{dwell}$, but the electrons are exposed to additional traps.

Accounting practices for traps between pixels varies throughout the literature.
\cite{bristow03im} and \cite{rhodes10} included the ability to step electrons through this region, and explicitly encounter new traps.
\cite{m10b} and \cite{a10} assumed trapping is instantaneous, so account for these traps by enlarging $V_\mathrm{pix}$ to include the region (only the degenerate combination $\rho_\mathrm{t}V_\mathrm{pix}$ can be measured).
\cite{short13} assuming trapping is not instantaneous, and the transfer time is so short that no electrons will be captured.
In practice, this decision will only matter at the image level if the clock sequence changes or, if trapping is not instantaneous, by influencing the initialisation of traps (traps under integration phases become filled during exposure, or the first operation of figure~\ref{fig:flowchart}, while traps under barrier phases remain empty; Hall et al.\ in prep.).

\subsubsection{Accounting for multiple transfers at once} \label{sec:balance}

Early codes \citep[e.g.\ ][]{bristow03im,rhodes10,m10a} computed the effect of every pixel-to-pixel transfer individually.
In a CCD with $n_x\times n_y$ pixels, electrons in the row of pixels closest to the serial register take only one parallel transfer to be read out, but electrons in the farthest row much undergo $n_y$ parallel transfers.
Full readout requires $n_xn_y(n_y+1)/2$ parallel transfers, which is more than $10^9$ for {\it ACS}/{\sl WFC} ($n_y$=2048 in {\sl ACS}/{\sl WFC} and {\it Euclid} or 4500 in {\it Gaia}).
This is fast in hardware, but performing this many iterations of the inner loop (boxes 3--5) in figure~\ref{fig:flowchart} takes a long time to reproduce in software.

\vspace{4mm}

To speed up their software, \citet{short13} and \citet{a10} independently introduced the same approximation.
If $\rho_e\gg\rho_\mathrm{t}$, the size of the electron cloud does not change dramatically from its first transfer to its last.
If the density of charge traps in each pixel is (modelled as) constant, the population of exposed charge traps is also the same at each transfer.
Every transfer is thus identical.

When the above assumptions hold, we need compute only the first pixel-to-pixel transfer (which shifts electrons by one pixel), and multiply its effect by the number of transfers.
In every column of the {\it ACS}/{\sl WFC} CCDs, the number of transfers faced by electrons in pixel rows 1 to 2048 is simply the vector 
\be
y_1=y=[1, 2, 3, ...\ 511, 512, 513, ...\ 2046, 2047,2048]
\ee
i.e.\ electrons in the first row of pixels take one step straight into the serial register, electrons in the row above take two steps... and the farthest electrons take 2048 steps.
Multiplying the effect of one transfer (with suitable error catching) is algorithmically equivalent to multiplying the number of electrons captured in one step, and this is the origin of the factor $y$ in equations~\eqref{eqn:m10vol}, \eqref{eqn:densdriven1} and \eqref{eqn:densdriven2}.
We shall call this the `express' $E=1$ approximation, because the effect of each pixel-to-pixel transfer has been computed only once. 
Reading out the entire CCD requires only $n_xn_y$ iterations of the inner loop in figure~\ref{fig:flowchart}.

\vspace{4mm}

\citet{a10} suggested a generalisation for better accuracy, which \citet{m10b} implemented.
In reality, not all pixel-to-pixel transfers are identical. 
The assumption $\rho_e\gg\rho_\mathrm{t}$ breaks down if $\rho_\mathrm{t}$ is large (e.g.\ in a severely damaged CCD) or $\rho_e$ is small (e.g.\ in dark frames, short exposures or at wavelengths where the sky is very faint -- see section~\ref{sec:holger}).
In this situation, once a cloud of electrons has undergone all transfers but one, the size of the charge packet may have changed significantly.
If the charge packet is smaller (or larger) during the final transfer, it will be exposed to fewer (or more) traps than during the first.

We can account for this evolution by applying the first transfer only a limited number of times, then recalculating the effect of subsequent transfers.
As a concrete example, consider a case in which we apply the first transfer up to only 512 times.
We dub this the $E= 2048/512=4$ approximation, because the effect of some transfers will need to be recomputed 4\ times.
The effect of the first 512 transfers can be computed by performing one transfer as before, but capping $y_{4,1}=\mathrm{max}\{y,512\}$.
The effect of the next 512 transfers is then recalculated, using the updated size of each charge packet.
The procedure is then repeated for a third and a fourth time.
In each case, the number of captured electrons is multiplied by one row of 

\begin{widetext}
\begin{equation*}
y_4=\left[
\begin{array}{rrrrrrrrrrrrrrrrrrrrrrrr}
1 & 2 & 3 & ... & 511 & 512 & 512 & 512 & ... & 512 & 512 & 512 & 512 & ... & 512 & 512 & 512 & 512 & ... & 512 & 512 & 512\\
0 & 0 & 0 & ... & 0 & 0 & 1 & 2 & ... & 511 & 512 & 512 & 512 & ... & 512 & 512 & 512 & 512 & ... & 512 & 512 & 512\\
0 & 0 & 0 & ... & 0 & 0 & 0 & 0 & ... & 0 & 0 & 1 & 2 & ... & 511 & 512 & 512 & 512 & ... & 512 & 512 & 512\\
0 & 0 & 0 & ... & 0 & 0 & 0 & 0 & ... & 0 & 0 & 0 & 0 & ... & 0 & 0 & 1 & 2 & ...  & 510 & 511 & 512\\
\end{array}\right].
\end{equation*}
\end{widetext}
For the first 512 pixels, only 1 transfer is ever calculated; for pixels 513-1024 there are 2 calculations; for pixels 1025-1536 there are 3 calculations; and for pixels 1537-2048, the maximum number of transfers explicitly calculated is 4.
It can be checked that each pixel is acted upon by the correct number of transfers because the columns of $y_4$ add up to the same, monotonically increasing values as $y_1$.

The above approach can be generalised to an approximation with any value $E$ between 1 and 2048 (note that we also refer later to $P\equiv n_y/E$). 
In the limit of $E=2048$ (or $P=1$), this recovers the slow algorithms of \citet{bristow03im} and \citet{m10a} in which the effect of every transfer is computed afresh.
For this, $y_{2048}$ is a square, upper-triangular matrix containing only ones and zeros.
Since later transfers are not applied to many pixels (the zero entries in $y_E$), they do not even need to be computed for all pixels.
Counting the non-zero entries in $y_E$, an efficient implementation will require $n_xn_y(E+1)/2$ iterations of the inner loop in figure~\ref{fig:flowchart}.

\subsubsection{Release of electrons from charge traps (box 5)} \label{sec:release}

After a short delay inside a charge trap (the length of which depends on the trap species), captured electrons are eventually released into whichever charge packet is then nearby.
The \citet{m10b} and \citet{short13} algorithms model the probabilistic release of electrons as an exponential decay.
Each species of charge trap (at least 3 species in {\sl HST}, and 4 in {\it Gaia}) has a characteristic release profile with a different half-life.
Releasing a constant fraction of electrons at each time step is computationally easy.
More importantly, the half-lives changed in a predictable way when {\sl HST}'s operating temperature was lowered in 2006 (for more details see figure~\ref{fig:trap_density} of \citealt{m10a}), and behave as expected within a wide range of operating conditions for {\it Gaia} \citep{prod12,short13}.

The \citet{a10} algorithm instead releases traps according to a function described by a look-up table, which depends upon the time since the electrons were captured \citep[and thereby reproduces a 3-exponential decay,][]{m10b}.
Only one computational species of trap is required to produce any trail profile.
There is extra overhead to keep track of the time but is still faster than computing the trails for 4 separate trap species.

\citet{a10} also found evidence for two different sets of trail profiles in {\sl HST} data: electrons near the bottom of a pixel (smaller packets of electrons than those considered by \citealt{m10b} or in this work) saw only fast-release traps, and electrons near the top saw additional slow release traps.
They suggested (see their \S6.2) that this discrepancy might disappear when $E>1$ algorithms are implemented.
However, if the effect remains, an elegant solution in the regime when the number of exposed traps is greater than the number of electrons might be a density-driven charge capture model in which the fast traps are filled first, and the slow ones are filled more slowly.

Various algorithmic tricks can be used to increase speed, for example by considering (for electron release) only traps that have previously been exposed to electrons.
Monitoring a high water mark of electrons against the fixed grid of traps, and considering only traps below this mark, offers a large speedup until a saturated pixel is encountered.
It is also possible to lower the water mark as the electron content in the traps approaches zero, and thus regain speed. 

\subsubsection{Loop over columns (box 7)} \label{sec:gpu}

This loop is trivially parallel, and may thus be ideal case for massive parallelization techniques that exploit either GPUs or MPI on multiple CPUs.

\subsection{Removing CTI trailing} \label{sec:algorithm_subtract}

The above approach describes a forward operation to {\em add} CTI trailing.
It is a software simulation that approximates the effect of CTI in radiation damaged hardware.
What we really need is a software algorithm to {\em remove} CTI trailing by reversing the readout.
Finding such an algorithm is a classic `inverse problem', and two generic approaches exist.

\subsubsection{Forward modelling}

The data analysis pipeline for {\it Gaia} \citep[e.g.][]{prod12,sea12} generates a model of every source and applies the \citet{short13} forward CTI transform, before matching the trailed model to the (trailed) data.
This approach requires accurately simulated data, which must be passed through the entire data analysis pipeline.
Since {\it Gaia} observes mainly stars, the simulations (of just a PSF) have the potential to be sufficiently accurate. 
Observations of galaxies and extended sources are harder to simulate accurately.
For limited applications, such as weak lensing with {\it Euclid}, it may be possible to simulate galaxies with sufficient detail for a similar forward-modelling approach \citep[e.g.][]{lensfit3,im3shape}. 

One unique ability of a forward approach is to model cosmic rays that hit the CCD during readout.
Electrons create by these cosmic ray hits undergo a smaller number of transfers than other electrons, become less trailed, and offer less shielding to downstream sources than would otherwise be expected.
Correcting their trailing probably requires that exact sequence of events to be followed precisely.

\begin{figure}
\begin{center}
\includegraphics[width=70mm]{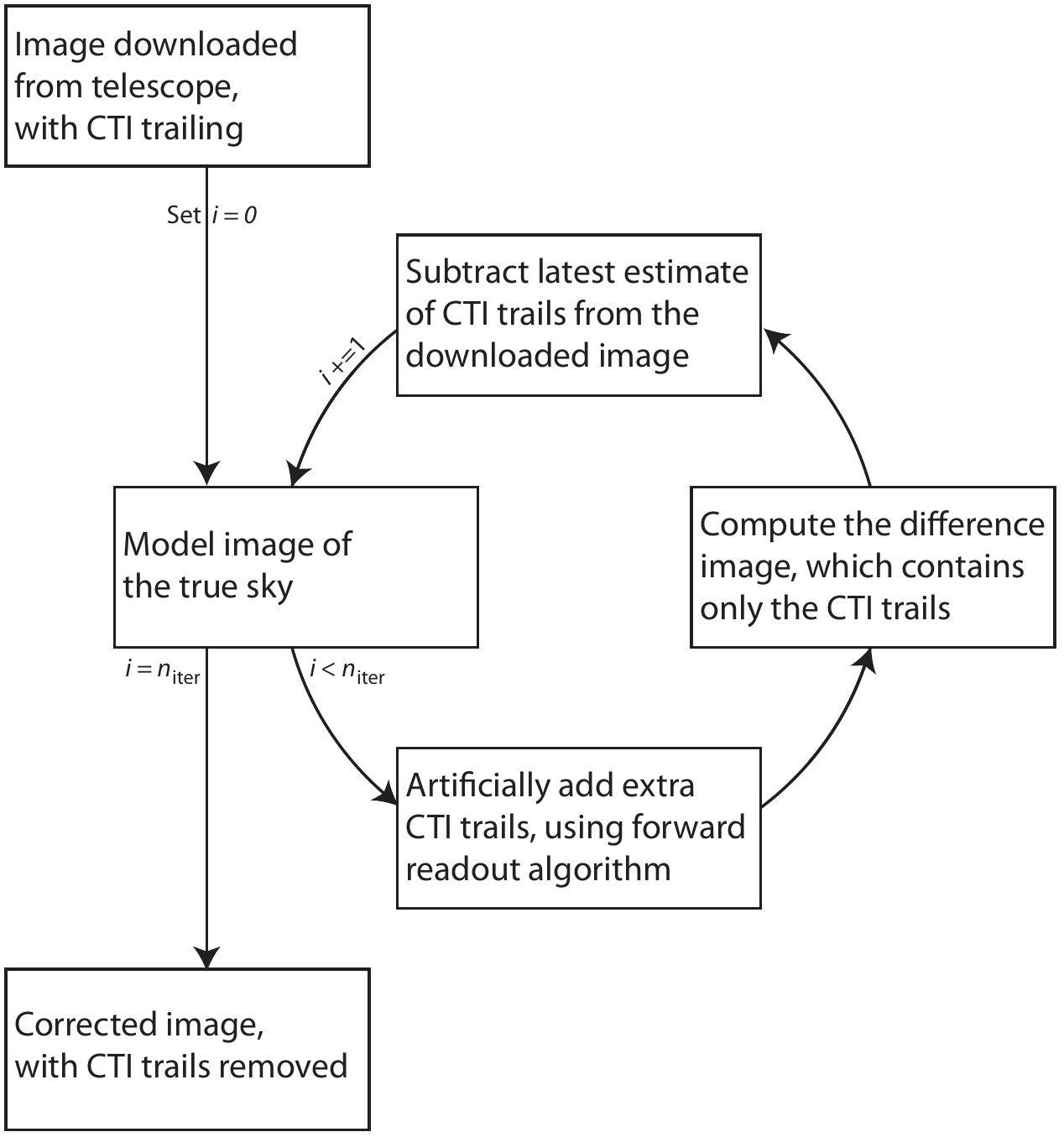}
\end{center}
\caption{Inverse operation to remove and thus correct CTI trailing.
This works by repeatedly applying the forward algorithm to add CTI trailing (figure~\ref{fig:flowchart}, shown here as the bottom box in the loop).
It converges towards an image that, when read out again, reproduces the (trailed) image downloaded from the spacecraft.}
\label{fig:flowchart_inverse}
\end{figure}

\subsubsection{Iterative inversion}

For more general workhorse applications, a flexible and standalone image correction scheme is better.
Trailing is (almost) the last process to happen during data acquisition, so it should ideally be corrected during the first stage of data analysis (after only bias subtraction and gain correction).
The \cite{m10a} and \cite{a10} pipelines developed for {\sl HST}/{\sl ACS} use an iterative method first suggested by \citealt{bristow02} and illustrated in figure~\ref{fig:flowchart_inverse} (for a more mathematical treatment, see table~1 of \citealt{m10a}).
This method iterates towards a (corrected) version of the image that, when trailed a final time, matches the (trailed) data.
Convergence requires $n_\mathrm{iter}$=5 \citep{a10} or 3 \citep{m10b} iterations, which correspondingly lengthens runtime -- but once it is run, a detailed image is available for all subsequent analysis and CTI can be forgotten.

\subsubsection{Dealing with read noise}\label{sec:huff}

The forward-modelling approach requires models of the true sky, which can be very accurately trailed because that model can be created without noise.
On the other hand, any iterative CTI correction will be fundamentally limited by the presence of read noise \citep{a10,cropper13}. 
Read noise is added after charge transfer so does not get trailed during readout, but it is incorrectly {\em untrailed} with the rest of the image during post-processing.
As evidenced by convergence to the wrong answer in even the idealised test of figure~\ref{fig:backward_conv}, the spurious untrailing of read noise introduces errors in measurements of photometry, astrometry and morphology.
Furthermore, spuriously untrailing the read noise increases its apparent rms, and thus increases the noise level noise in the final image (CTI correction is similar to an unblurring/sharpening operation, in that it increases the autocorrelation of pixels while introducing an anti-correlation of adjacent pixels).

\citet{a10} attempt to find the `minimum' CTI correction consistent with knowledge of the image and the read noise.
They apply a high-/low-pass filter to the image (see their \S 5.1), which isolates a component that is considered to be read noise, and running the correction on only the smooth component.
This certainly achieves a smoother final product, but in an unsatisfactory manner as e.g.\ the choice of filter threshold is ad hoc, and (high frequency) shot noise in the image {\em was} trailed.

We propose a different solution, based around the idea that the incorrect measurements after correction are purely caused by the presence of noise that is (anti-)correlated between pixels.
In similar situations, \citet{k2k}, \citet{bj02} and \citet{huff11} have noted that noise correlated more in the $y$ direction than the $x$ direction can lead to biases in centroids in the $y$ direction, plus a larger uncertainty in the centroid in the $y$ direction, which in turn creates biases in galaxy shapes.
Furthermore, this correlated noise can also produce selection biases, whereby more objects aligned in the $y$ direction are detected above a S/N threshold.
The three papers above developed a method to adjust the noise covariance matrix, by {\it adding extra random noise} that has then been correlated between adjacent pixels in exactly the right way to make the total noise white.
The decision to add noise in software is somewhat analogous the decision to implement a pre-/post-flash illumination in hardware.
For the cost of a small amount of extra noise, the data products become easier to handle for precision measurements of photometry, astrometry and morphology.
We shall test whether this also improves the CTI correction.

\begin{figure*}
\begin{center}
\includegraphics[width=180mm]{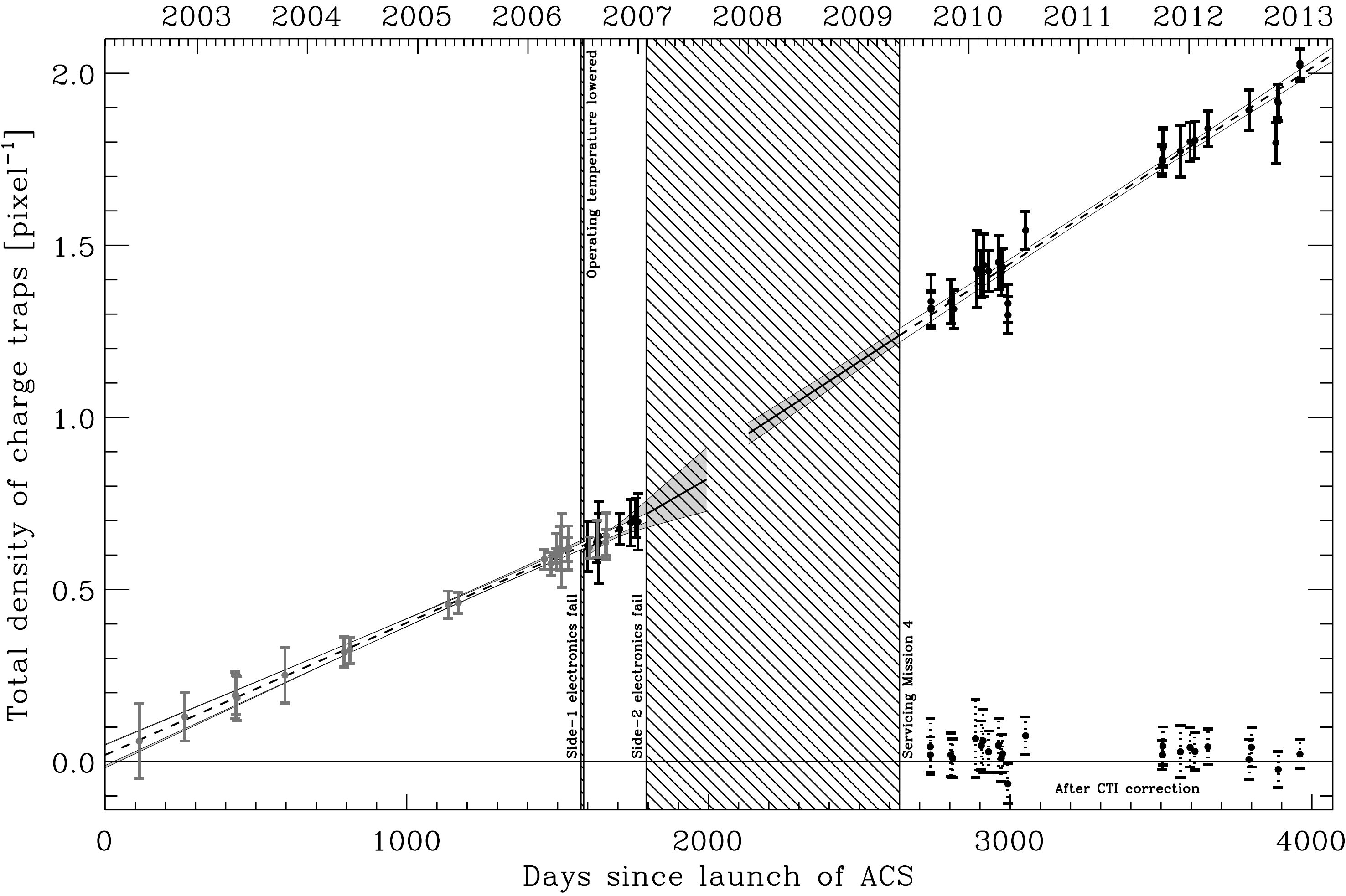}
\end{center}
\caption{Density of charge traps in the {\sl HST} {\it Advanced Camera for Surveys}/{\it Wide Field Channel}. 
Points with solid error bars show trap densities measured from the amount of trailing behind warm pixels (grey and black are different gain settings).
Points with dashed error bars show measurements behind the same warm pixels after correction with $n_\mathrm{iter}=6$ and other algorithmic choices as described in the text.
Hatched regions show significant periods when {\sl ACS} was offline.
}
\label{fig:trap_density}
\end{figure*}

To determine how much noise to add, we first identify a region of blank sky and measure the noise power spectrum $P_N(\mathbf{k})$ (in practice, we use a truly blank image containing only simulated sky background passed through the forward readout process, plus read noise).
We then follow the method described in section~4.2 of \citet{huff11} to construct an additional amount of synthetic noise such that the covariance of the total noise is 4-way symmetric.
In addition, we also try adding synthetic noise that makes the total noise white (zero covariance between pixels).
This uses a very similar method and just requires a larger rms.
We shall test in section~\ref{sec:newmodel} how much noise each version adds to a typical image.

\section{Calibration on real data} \label{sec:newmodel}

We shall now construct an up-to-date model of CCD detector readout for the {\sl HST} {\sl ACS}.
Using a combination of real, in-orbit data and simulations, we shall attempt to construct a code in which all approximations lead to sub-1\% inaccuracies.
Within this limit, we shall attempt to maximise computation speed.

\subsection{Measurements from real data} \label{sec:trap_species}

We measure CTI trails behind warm pixels in long F814W exposures taken as part of the CANDELS survey (GO-12444) in January 2013.
Warm pixels are useful because they should be perfect delta functions in the absence of CTI.
The observed trails have the same profile shape as those seen in \citet{a10}.
After accounting for `self-CTI' (multiple captures of an electron during readout, so that trail becomes longer than an exponential, in a way that depends on the background level and capture model), \citet{m10b} demonstrated that this profile is well-fit by a model in which three trap species have exponential release profiles, 
\be \label{eqn:tauchoice}
\tau=\{0.74, 7.7, 37\}~\mathrm{pixels}
\ee
and relative densities
\be \label{eqn:rhochoice}
\frac{\rho_\mathrm{t}}{\sum\rho_\mathrm{t}}=\{0.17, 0.45, 0.38\}.
\ee

However, recent trails have increased in {\it amplitude} -- reflecting an increased trap density.
From the {\sl HST} data archive, we identify long (1/3- or 1/4-orbit) exposures in F606W or F814W of uncrowded, extragalactic fields obtained between mid 2011 and early 2013.
These images have sky backgrounds $\simgt$50 electrons.
A fit to the inferred trap densities during this period is shown in figure~\ref{fig:trap_density}.
We have not updated our trap model for data obtained before Servicing Mission 4, because the lower trap density then makes the existing correction sufficient.

We introduce two improvements when fitting the trap densities.
In \citet{m10b}, we measured the trail amplitudes throughout an image, then fitted the global trap density, weighting all the trails equally.
In our improved procedure, we now recognise warm pixels far from the readout register are trailed most, so we now give measurements of their trails an appropriately higher weight during fitting than those near the readout register.
We also recognise that in these pixels, which have undertaken many transfers through a harshly degraded CCD, not all of the transfers will have been identical
(equivalent to the $E>1$ readout algorithm described in section~\ref{sec:balance}).
To account for this (and self-CTI), we pass a hot pixel through the full readout process (this can be done fast and analytically for a delta function), and iterate its flux until it contains $n_\mathrm{warm}$ after readout.
Both changes result in an increase in reported trap densities: approximately  7\% from upweighting pixels farther from the readout register and a further 1\% by accounting for the changing charge cloud size, although these numbers depend mildly on the number of electrons in a pixel.

Fitting the amplitude of trails for warm-hot pixels now requires a well filling model with $\beta$=0.478 and full well depth $w$=84700\,electrons.
Given this, the observed density of traps after Servicing Mission 4 is 
\be \label{eqn:rhott}
\rho_\mathrm{t}V_\mathrm{pix}=(1.66\pm0.01)+(5.65{\pm}0.24)\times10^{-4}(t-3374)\,\mathrm{pixel}^{-1},
\ee
where $t$ is the number of days since launch (see figure~\ref{fig:trap_density}).
The rate of degradation from radiation damage increased during the 2007--2011 Solar minimum, and is showing no sign of slowing.

\subsection{Tests on simulated data} \label{sec:simulations}

The challenge with real data is that we see the sky only after the effect of CTI, and we rarely know its truth.
Here we construct simulated data -- which we degrade with a readout model, then correct using the same readout model.
To test the internal stability of (the inversion of) the algorithm, we can explicitly compare corrected data to its appearance in the absence of CTI.
Note, however, that any inconsistency will not account for error on the model, such as the density or positions of charge traps.

\begin{figure*}
\begin{center}
\includegraphics[width=170mm]{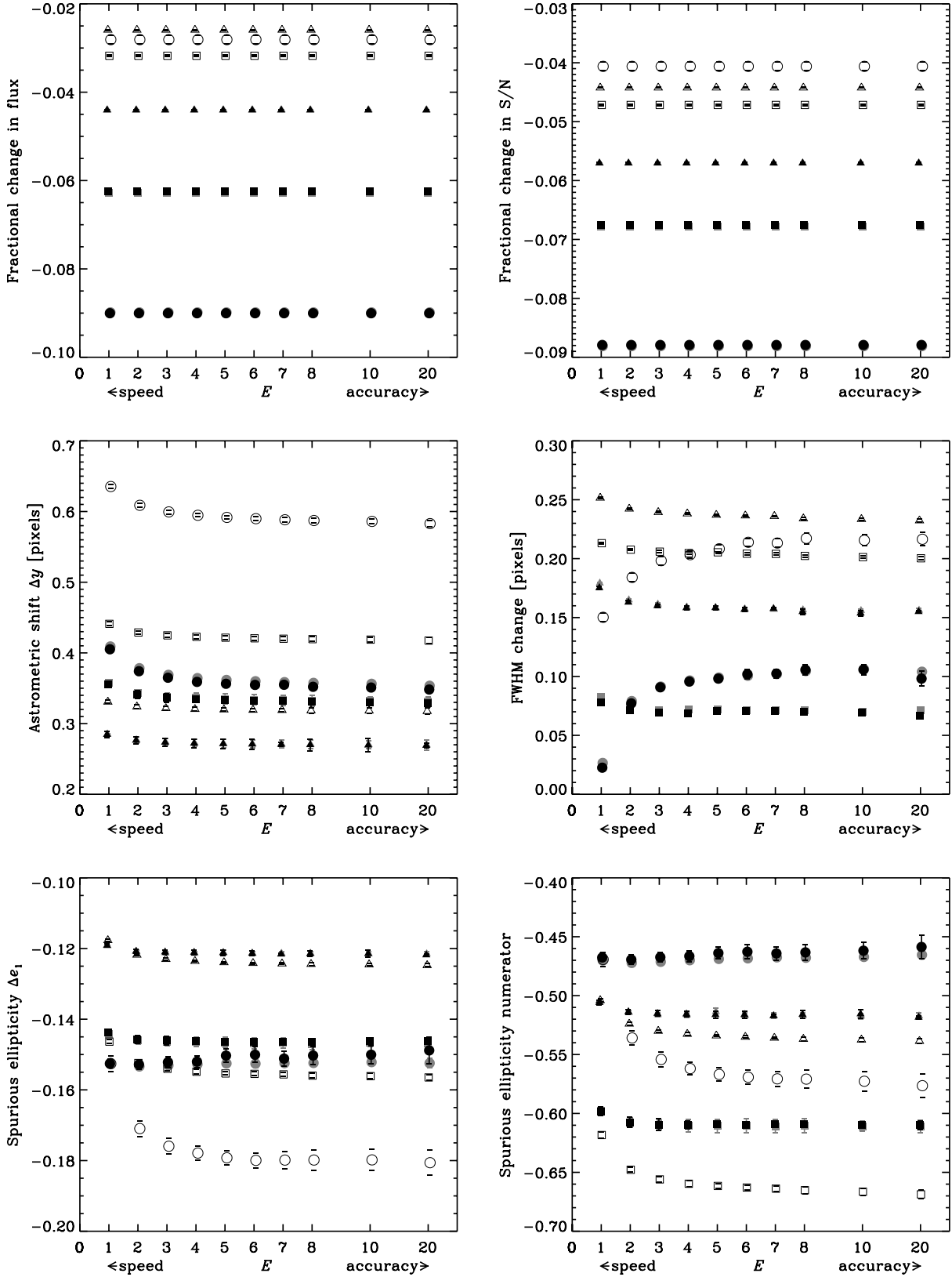}
\end{center}
\caption{Convergence of our (forward) CCD readout model.
The express parameter $E$ governs a tradeoff between speed ($E=1$, the result of a single pixel-to-pixel transfer is applied identically many times) and full realism ($E\rightarrow2048$, the effect of each transfer is calculated afresh). 
Circles, squares and triangles respectively show the effect on a galaxy with S/N=10, 50, 100.
Black, grey and white symbols show outputs of algorithms identical except that $n_\mathrm{levels}$=10000, 2048$V_\mathrm{pix}$max$(\rho_\mathrm{t})$, 1:  i.e.\ mimicking \citealt{m10b}, \citealt{a10} and \citealt{short13}.
The different-sized errors mainly reflect the different number of galaxies simulated in each case.
}
\label{fig:forward_conv}
\end{figure*}

\begin{figure*}
\begin{center}
\includegraphics[width=170mm]{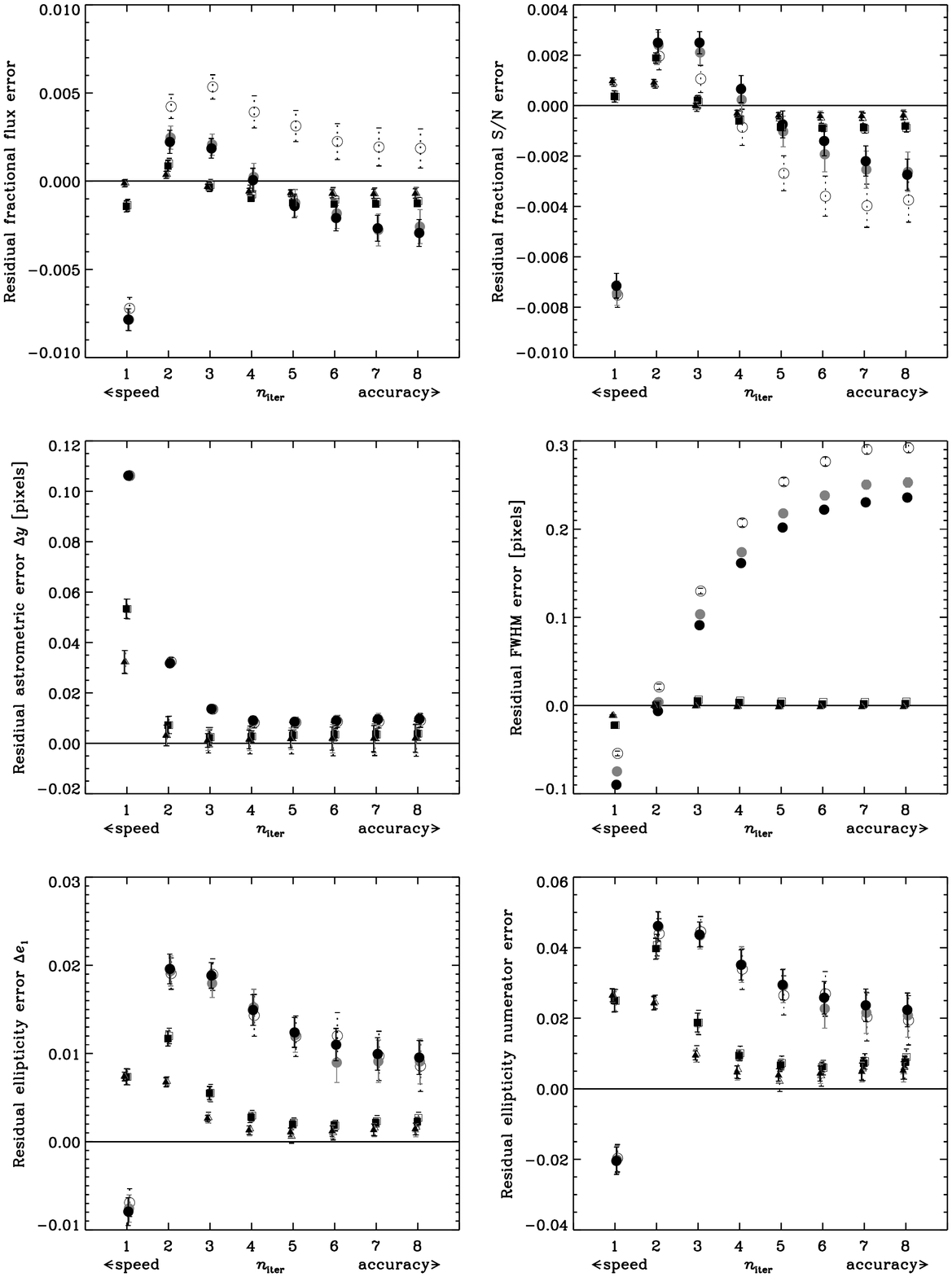}
\end{center}
\caption{Convergence of our (inverse) CTI-removal algorithm, after some number $n_{\rm iter}$ of iterations.
The same forward algorithm is used to add then remove CTI trailing.
In the absence of read noise, correction is perfect (this case is not shown); with read noise, the iteration converges to the wrong image.
Circles, squares and triangles respectively show the miscorrection of a galaxy with S/N=10, 50, 100.
Black, grey and white symbols show results after a raw CTI correction, noise whitening to make total noise 4-way symmetric, and noise whitening to make the noise completely white.}
\label{fig:backward_conv}
\end{figure*}

\subsubsection{Tests on simulated science data}

To investigate the effect of CTI on complex, extended sources, we simulate images of faint galaxies seen by \acswfc.
We simulate several hundred thousand noise realisations of a circular, exponential, just-resolved (scale size $0.06\arcsec$) galaxy, convolved with a {\sc TinyTim} \citep{tinytim1,tinytim2,tinytim3} model of the {\sl HST} point spread function (PSF) at 814nm, centered randomly within a pixel at position $y=2048$ from the serial register.
We place this galaxy on a $\sim$$100$ electron (and noisy) sky background, simulate parallel CCD readout, then add Gaussian read noise with $4.0$ electron rms per pixel.
We scale the flux of the galaxy and repeat this analysis for a galaxy detected by {\sc Source Extractor} \citep{sex} at a mean signal to noise ratio of S/N=10, 50, and 100.
In these cases, {\sc Source Extractor} measures a mean {\sc FWHM} size of 2.48, 2.00, 2.00 pixels respectively.

We use {\sc Source Extractor} to measure the total flux ({\sc FLUX\_AUTO}), signal-to-noise ratio S/N ({\sc FLUX\_AUTO}/ {\sc FLUXERR\_AUTO}) and size ({\sc FWHM\_IMAGE})  of the image $I(x,y)$, and {\sc RRG} \citep{rrg} to measure its $y$ position and PSF-convolved ellipticity
\be
e_1=\frac{\int I(r,\theta) ~ W(r) ~ r^2{\mathrm cos}(2\theta) ~ r {\mathrm d}r{\mathrm d}\theta }{\int I(r,\theta) ~ W(r) ~ r^2 ~ r {\mathrm d}r{\mathrm d}\theta} \label{eqn:rrge1}
\ee
where the weight function $W(r)=\mathrm{exp}(-r^2/2\sigma^2)$ is centered such that the first moments are zero and $\sigma$, which is fitted to each noisy image, is $\sim$3.3, 4.4, and 4.6~pixels on average for the S/N=10, 50, and 100 galaxies.
We also record the numerator of $e_1$, which is dimensionful but linear in image flux and therefore avoids potential noise-induced shape measurement biases \citep{shapelets4,zhang11,refregier12,melchior12}.

We first measure all these quantities on a version of the simulated image without any CTI, and record their `true' values.
We then remeasure the quantities on versions of the simulated images in the presence of CTI, using our {\sl ACS}/{\em WFC} readout model (assuming an observation date of 1~January 2013, 3959 days since the launch of {\sl ACS} and requiring a total of 1.989 traps per pixel).
The effect of forward CTI trailing is shown in figure~\ref{fig:forward_conv}, for various algorithmic choices.

Finally, we apply our CTI correction algorithm to the degraded simulated images, and remeasure the quantities.
The net effect of CTI trailing after correction is shown in figure~\ref{fig:backward_conv}, for various algorithmic choices.

\subsubsection{Tests on simulated dark (low signal) exposures} \label{sec:holger}

Accounting for trap occupancy in a finite number of $n_\mathrm{levels}$ (section~\ref{sec:nlevels}), and accounting for multiple transfers simultaneously via the `express' $E$ approximation (section~\ref{sec:balance}), relied on the condition that $\rho_e\gg\rho_\mathrm{t}$.
This assumption breaks down if sufficient radiation damage created a high trap density, or if the signal and background are low (e.g.\ in dark exposures or at wavelengths with faint emission and sensitivity such as with {\sl WFC3}/{\sl UVIS} data).
In terms of CTI correction, the low-signal regime is defined in terms of the {\it total} number of photoelectrons, because charge transfer does not discriminate between source counts from astronomical objects or the sky background. 
High sky backgrounds pre-fill traps that do not then affect electrons, while a low sky background makes images more susceptible to CTI. 

\begin{figure}
\begin{flushright}
\includegraphics[width=74mm]{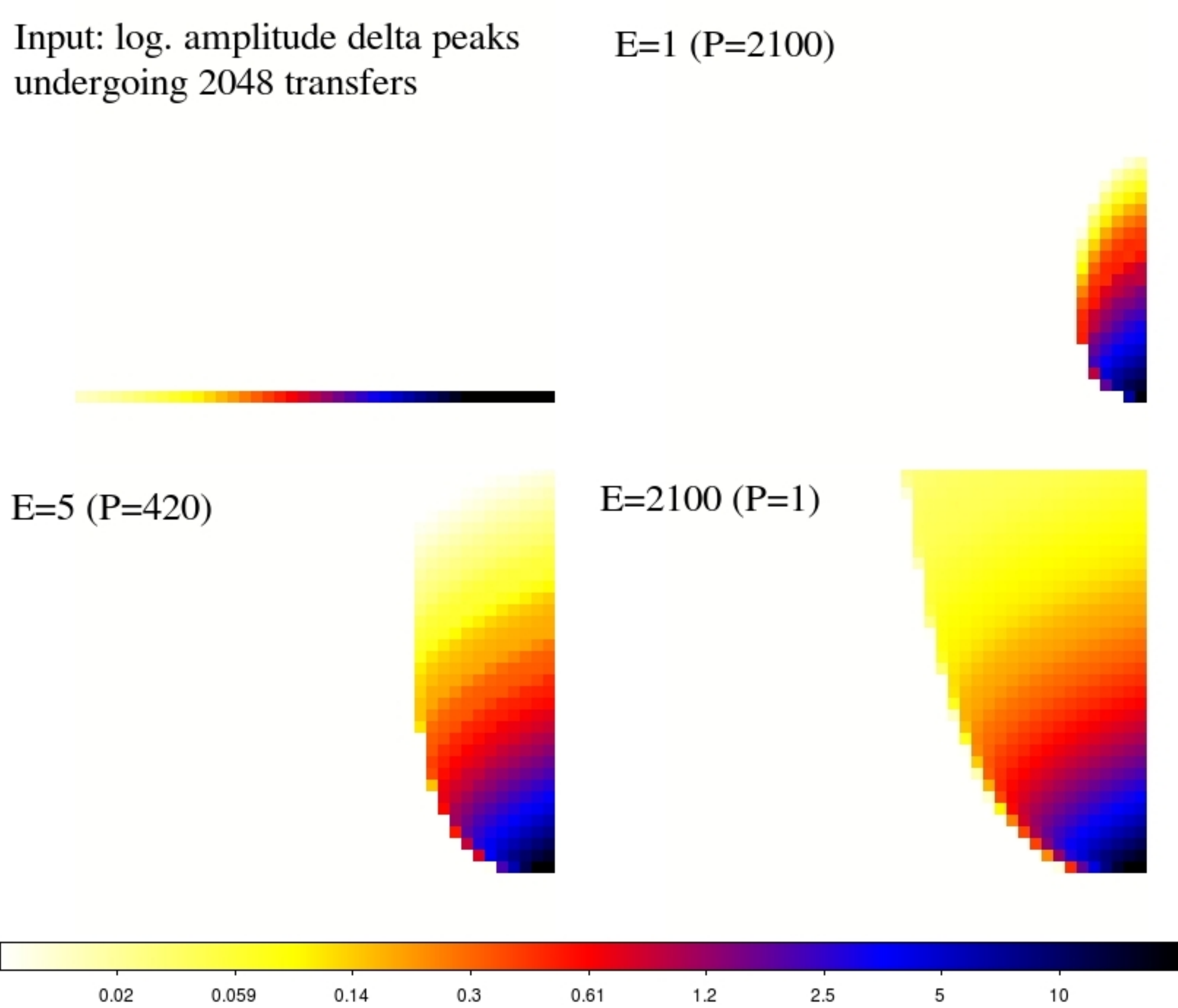}\\~\\
\end{flushright}
\begin{center}
\includegraphics[width=76mm]{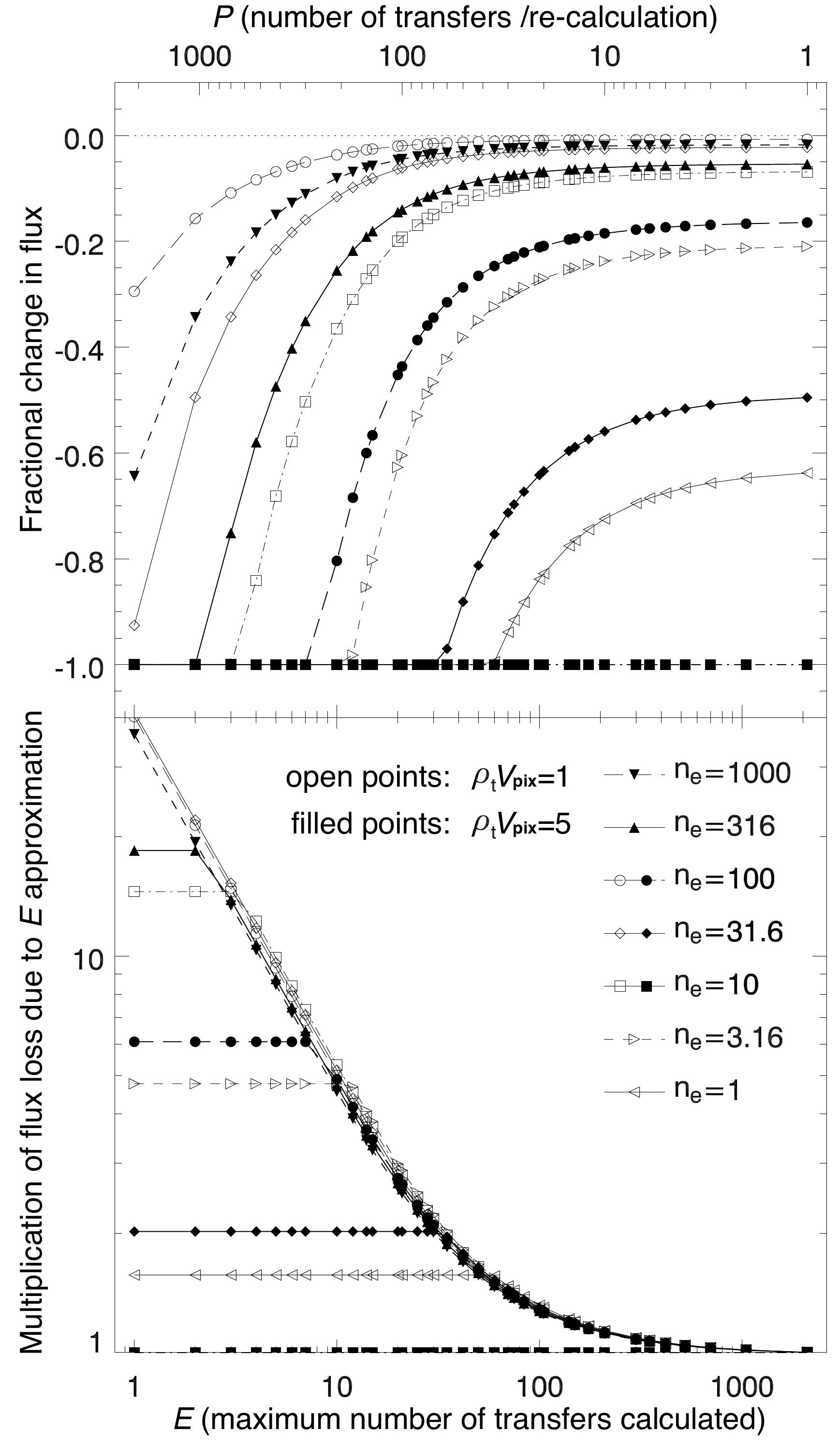}
\end{center}
\caption{In the low signal-to-trap regime, approximations made to speed up processing have a greater impact on performance.
\emph{Top panel:} Test data showing a ramp of hot pixels before and after CTI trailing (with readout towards the bottom of the page), for different choices of $E$. 
The colour scale is logarithmic.
\emph{Middle panel:} Fractional charge loss from the image as a function of $E$.
Open (filled) symbols denote different input signal levels and a trap density
$\rho_{\mathrm{t}}V_\mathrm{pix}\!=\!1\,\mbox{pixel}^{-1}$ ($5\,\mbox{pixel}^{-1}$), for $\tau\!=\!2$\,pixels.
\emph{Bottom panel:} Excess charge loss due to the express approximation, compared to the `natural' flux loss with slow-but-steady $E=n_y$.} \label{fig:lowsig}
\end{figure}

To investigate the behaviour of our algorithm in the low signal regime, we construct test data consisting of delta function peaks of varying heights (similar to the hot pixels used in section~\ref{sec:trap_species}), 2048 pixels from the serial register and on zero background. 
We then apply an artificial readout, through a high density of traps that all have $\tau=1$\,pixel.
Note that all the traps we use for this test are of the same species in order to avoid the complicating ambiguity of which to fill first.
The upper panel of figure~\ref{fig:lowsig} shows an example of our test data on input (a ramp in the $x$ direction or a series of delta functions in the $y$ direction), and after simulated readout with three different choices of $E$.

Since $\rho_\mathrm{t}\gg\rho_e$ for these tests, one might expect that some of the electrons will be captured and not released until after readout.
Such electrons would effectively be lost from the image.
However, the parabolic shape of the trailed regions in figure~\ref{fig:lowsig} reveals an interesting aspect of any volume-driven CTI model -- even in the case of best accuracy ($E$=$n_y$ or $P$=$n_y/E$=1), in which the effect of each pixel-to-pixel transfer is calculated afresh.
During the first transfer, traps indeed capture all the electrons up to a certain height.
Depending on the trap species' characteristic release time $\tau$, the electrons are gradually released.
The electrons fall to the bottom of the potential well where, because of the high trap density, they are immediately recaptured.
For initially small charge clouds, the electrons never emerge from the traps back into the image (they are transported off the edge of the CCD).
With larger charge clouds, however, the traps at the bottom of the potential eventually become saturated.
Electrons released from traps at the top of the potential fall to the bottom and can no longer be recaptured. 
These electrons finally re-emerge back into the image.

To quantify the flux loss from an image, we sum over the CTI trail downstream of the original peak for 52 pixels, and compute the fractional charge loss from the image before and after readout, $\delta\equiv(\sum n_e^\mathrm{trailed}-\sum n_e^\mathrm{true})/\sum n_e^\mathrm{true}$.
We find that fast and intermediate trap species ($\tau$ from a few tenths to a few pixels) rapidly deplete the signal whenever the first trap captures a significant fraction of electrons.
After 2048~transfers, the signal loss from the image is
\be
\delta\approx-8.46\!\times\!10^{-4}\left(\frac{n_e}{w}\right)^{-0.647}\left(\rho_\mathrm{t}V_\mathrm{pix}\right)^{1.221}\,\tau^{0.114},
\ee
assuming $\beta$=0.478 and full well depth $w$=84700 electrons (errors on the exponents are less than 0.03).
Slow trap species deplete the signal even more severely. 
For example, traps with $\tau=40$\,pixels remove $20$\% of a signal from an image in the same conditions ($n_e$ and $\rho_\mathrm{t}$) where traps with $\tau$=1\,pixel remove only $1$\%.

Unfortunately, we find that additional flux can be artificially lost when approximations are made in the readout algorithm.
In particular, the timing of the electrons' re-emergence into the image depends upon the values of $n_\mathrm{levels}$ and $E$.
The $n_\mathrm{levels}$ parameter affects this because, although our algorithm is globally volume-driven, it is set up to be density-driven inside each level. 
Low values of $n_\mathrm{levels}$ increase the number of traps in the bottom level, delay trap saturation, and lengthen the time before any electrons re-emerge.
However, this effect is minor: lowering $n_{\mathrm{levels}}$ from $10,000$ to $2048$ only affects trails with $n_e$$<$1\,electron by more than a few percent.

Performance is more severely degraded by lowering $E$.
As described above, some flux is lost from an image even in the slow-but-accurate extreme $E=n_y$ (see the right-hand side of the middle panel in figure~\ref{fig:lowsig}).
As $E$ decreases, the effect of each transfer is updated only after $P\equiv n_y/E$ transfers, so the number of electrons captured in each but the first transfer is overestimated.
Furthermore (but less importantly), because more electrons are allowed to be captured during each computed transfer, the `bottom' level is effectively enlarged.
Both effects act to delay the eventual re-emergence of trapped electrons back into the image.
We find that there is a minimum $E$ at which a cloud initially containing $n_e$ electrons can be registered with a certain fraction of its flux for a given detector (defined by $\rho_\mathrm{t}V_\mathrm{pix}$ and $\tau$).
For $n_e$=100, the most accurate output ($E=n_y$) is reproduced to within $10\%$ if $E>5$, or within $1\%$ if $E>20$.

Conveniently, the factor by which flux loss is artificially enhanced by the `express' approximation is nearly constant with respect to $n_e$ and $\rho_\mathrm{t}$ (i.e.\ it depends only on $E$).
The bottom panel of figure~\ref{fig:lowsig} replots data from the middle panel, renormalised to the `natural' performance with $E=n_y$.
While `natural' flux loss from the image cannot be avoided, the ordinate quantifies additional flux loss for $E<n_y$, which can be avoided at the expense of computational speed.
The lined-up data points can be used to define a minimum suitable $E$ in the low-signal regime.
The multiplicative factors can be as large as 30 -- but consider that this is 30 times a very small amount.
Our overall goal of $<1\%$ residual from CTI applies to the electrons in a typical science image, rather than to the very few electrons in these simulations.

\section{Improved CCD readout model} \label{sec:choices}

We now have all the necessary data in hand to make suitable choices for the algorithmic options discussed in section~\ref{sec:algorithm}.
Our choices are summarised in the bottom row of table~\ref{tab:metamodel}.

\subsection{Adding CTI trailing}

Here we describe and justify our choices for the forward readout algorithm, in subsections mirroring those in section~\ref{sec:algorithm_add}.

\subsubsection{Initialising charge traps}

We initialise all traps as empty, assuming a long exposure during which electrons could escape.

\subsubsection{Capture of electrons into charge traps} \label{sec:properties_choice}

We use a volume-driven model of charge capture (equation~\ref{eqn:m10vol}) with $\beta$=0.478 and $w$=84700\,electrons.
We model the growth of total trap density via equation~\eqref{eqn:rhott}.
Supported by \cite{a10}'s empirical findings that all {\sl ACS}/{\sl WFC} traps that could capture an electron do so on their first opportunity, we assume electrons are instantaneously captured by any exposed traps.

\subsubsection{Monitoring trap occupancy} \label{sec:management_choice}

The value of $n_\mathrm{levels}$ required for suitable performance can also be learned from in figure~\ref{fig:forward_conv}.
The slow-but-sure limit of $n_\mathrm{levels}$=10,000, in which the accounting for trap occupancy is finely grained, is shown by black points.
We space our traps linearly in terms of the occupied pixel volume, as it is the captured traps that make a difference, rather than the free ones.
Using $n_\mathrm{levels}=2048\times$max$(\rho_\mathrm{t}V_\mathrm{pix})=1860$ boosts speed by a factor 3.7 while maintaining $<1\%$ accuracy, as shown by grey points.
Halving the number of traps further \citep[as tried by][and reproduced by us, but not shown on the plot for the sake of clarity]{m10b} does indeed introduce $\sim1\%$ errors.

Using $n_\mathrm{levels}$=1 \citep[as advocated by][but implemented within our code]{short13} is not sufficiently accurate for this application. 
The white points in figure~\ref{fig:forward_conv} show the consequences.
The model's approximations break down when the size of the electron cloud changes between adjacent pixels.
For example, the large electron cloud representing the central peak of a star image will fill many charge traps.
After that charge cloud has moved along, some of the traps will immediately release some of their electrons.
When the same traps see fewer electrons in the trailing wing of the star, the \citet{m10a} and \citet{a10} algorithms will allow some of those electrons to be captured.
However, the \citet{short13} algorithm will known only that many traps are still full, so capture no more free electrons\footnote{This limitation is present only {\em within} individual objects. The full {\it Gaia} data processing pipeline allows bright `up-stream' sources (or charge injection lines) to pre-fill and shield traps from `down-stream' faint sources by treating them as a variable background during (box~2) trap initialisation \citep{prod12,sea12}.}.

We use $n_\mathrm{species}=3$ and account independently for each species.
Bundling all three trap species together \citep[as advocated by][]{a10} results in a speed boost by a factor 1.6, at no loss of accuracy.
This is not as great as it could be because of the overhead from monitoring the time since they were filled (and emptying them accordingly).
There is also overhead in the complexity of this code, particularly when it comes to changing the operating temperature etc., so we choose to keep it simple.
However, if the instrument parameters were completely fixed, the freedom to arbitrarily adjust the shape of trail profiles has the potential to increase accuracy via a small number of extra parameters, and might be useful in the future.

\subsubsection{Pixel-to-pixel parallel transfers}

If charge capture is indeed instantaneous, we can also model parallel transfer as instantaneous, and the region between pixels can be generically considered to enlarge $V_\mathrm{pix}$.
Exactly how much it enlarges $V_\mathrm{pix}$ may depend on the shape of the electric potential in the CCD during transfer phases.
However, the only observable quantity is $\rho_\mathrm{t}V_\mathrm{pix}$, so we can merely fix $V_\mathrm{pix}$ and obtain an effective $\rho_\mathrm{t}$.

Note that some codes \citep{bristow03im,rhodes10} include a (fairly crude) implementation of parallel transfer through $n_\mathrm{phases}$ multiple phases.
Before running the algorithm, the image array is artificially expanded from $n_x$$\times$$n_y$ to $n_x$$\times$$n_\mathrm{phases}n_y$ (with each row of pixels interleaved with $n_\mathrm{phases}-1$ rows of empty pixels), and trap properties are redefined as $\rho_\mathrm{t}V_\mathrm{pix}/n_\mathrm{phases}$ and $\tau n_\mathrm{phases}$.
We have implemented this approximation and begun testing its consequences.
For a volume-driven model assuming instantaneous charge capture, we find no difference except that short traps effectively delay fewer electrons (and if the trail profiles are exponential, this decrement is degenerate with their effective density $\rho_\mathrm{t}$; Israel et al.\ in prep.).

\subsubsection{Accounting for multiple transfers at once} \label{sec:express}

We first test convergence of the forward CTI algorithm that adds trailing.
A crucial balance between speed and accuracy in the forward algorithm is set by the `express' $E$ parameter (see section~\ref{sec:balance}).
The change in measurements in the presence of CTI, compared to an ideal image, is shown in figure~\ref{fig:forward_conv}.
These converge as $E$ increases (not to zero, but to whatever value a long-winded CTI algorithm would predict).

In typical science data, to achieve $\sim$$1$\% precision in most observables within a minimum CPU runtime, we recommend using $E\ge5$.
In images with low $\rho_e/\rho_\mathrm{t}$ (especially low background), the artificial flux losses introduced by the express approximation can be mitigated by increasing $E$.
In dark exposures, we recommend that $E$ should be $\ge20$.
This slows runtime on the dark exposures by a factor 4.
However, this may not significantly affect {\it total} runtime, because (a) the number of dark exposures is typically much smaller than the number of science exposures, and (b) the `high watermark' method (section~\ref{sec:release}) greatly speeds execution when $n_e$ is low, counteracting this effect.

\subsubsection{Release of electrons from charge traps}

Since the CTI trails in {\sl HST} observations are still well-fit by a sum of exponentials, we keep the \citet{m10b} trap model with $n_\mathrm{species}=3$ species that have exponential release in characteristic times $\tau$=\{0.74, 7.7, 37\}\,pixels and relative densities \{0.17, 0.45, 0.38\}.
This parameterisation remains sufficiently flexible to easily model performance over the range of past operational temperatures.

\subsubsection{Loop over columns}

We tried recasting the algorithm to implement this loop via NVidia CUDA on a GPU.
In practice, optimizing the code for GPUs is limited by the large amount of data needed to be transferred to and from the internal memory in administrating the traps.
Also, our tests showed that double precision is indispensable in the calculations, so this limits the GPU hardware (using NVidia's CUDA) to expensive {\it Tesla} GPUs.
We therefore find that, at present, it is thus more economic to perform parallel processing on a large multi-core CPU system (either with openMP or just running simultaneous jobs) than on a GPU system.

In contrast, note that experiments at STScI found single point precision sufficient, and GPUs there have resulted in more than 100$\times$ speedups (J.\ Anderson, 2013, pers.\ comm.).

\subsection{Removing CTI trailing}

Here we describe and justify our choices for an iterative method to invert CCD readout and remove CTI trails at the pixel level.
Subsections mirror those in section~\ref{sec:algorithm_subtract}.

\subsubsection{Number of iterations for inversion} \label{sec:niter}

The iterative method (figure~\ref{fig:flowchart_inverse}) operates for some number $n_\mathrm{iter}$ of cycles.
Our tests on simulated images (black points in figure~\ref{fig:backward_conv}) show that $n_\mathrm{iter}\ge6$ iterations are required for all measurements to converge, especially with the very faint (S/N=10) galaxies.

In some situations, fewer iterations may be sufficient, or even preferred.
Measurements of brighter galaxies (S/N=50, 100) and low-order (e.g.\ astrometry) measurements of very faint galaxies have converged by $n_\mathrm{iter}=3$.
This takes only half the computation time.
Furthermore, each iteration amplifies the read noise.
The contribution of read noise to the autocorrelation function of test image pixels at $y$=2000 is ideally 4.0, but 5.03 after 3 iterations, and 5.15 after 6 iterations.
It may therefore be prudent to perform no more iterations than necessary.

Note that even with very many iterations, the algorithm does not converge to the desired solution (shown in each panel as a horizontal line), despite exactly the same algorithm being used to add and remove CTI trails.
If read noise is not added to the simulations, the convergence is perfect.
Two issues related to read noise cause the incorrect convergence.
First, the read noise is amplified during CTI correction and, at low S/N, {\sc Source Extractor} in not stable to background noise level.
Second, the read noise is spuriously {\it un}trailed during CTI correction, when it was never trailed to begin with.
This introduces biases in even the more robust RRG measurements of astrometry and shape, which we shall attempt to deal with in the next section.

\subsubsection{Dealing with read noise} \label{sec:readnoise}

We have verified that all the measurements converge to the correct answer (i.e.\ that in the absence of CTI) if we add CTI and immediately correct it using the same readout model.
However, if we add read noise between adding and removing CTI, the correction converges to the wrong answer.
If we blindly ignore the read noise, astrometric shifts for the faintest galaxies are under-corrected from about 0.36\,pixels to 0.01\,pixels (but not zero), and their ellipticity is over-corrected from $\langle e_1\rangle$=$-0.15$ to $+0.01$.
This latter overcorrection is roughly proportional to the amount of read noise squared (Israel \etal\ in prep.).

Our measurements of galaxy position and shape were obtained with RRG, which is carefully designed to be able to average away the effect of background noise from a large sample of galaxies.
These (and RRG measurements of flux, which we have not shown) are well-behaved during the correction, and even if the noise whitening process increases the amount of noise (see section~\ref{sec:huff}).
Indeed, these measurements are completely unaffected by the addition of this correlated noise.

Measurements with {\sc Source Extractor} (flux, S/N, FWHM in figure~\ref{fig:backward_conv}) of very faint galaxies (S/N=10) are not robust to the extra noise.
In this case, the noise whitening process has actually made the correction worse --- although the instability of the measurements suggests that this conclusion is random and that the correction was probably equally as likely to have got better.

We conclude from this that adding noise to an image after CTI correction does not improve any measurements, even if the noise is correlated in such a way to make the total noise white.

\section{Performance tests on real data} \label{sec:results}

The final proof that our choices are appropriate lies in the correct removal of charge trailing from real imaging.
We shall now correct a large back-catalogue of \acswfc\ data, then test the shapes of warm pixels (which should become delta-function spikes) and galaxies (which should be randomly oriented).

\subsection{Warm pixels}

Recall from section~\ref{sec:trap_species} that the density of charge traps is $\rho_\mathrm{t}V_\mathrm{pix}=(1.66\pm0.01)+(5.65{\pm}0.24)$$\times$10$^{-4}(t-3374)$\,pixel$^{-1}$, where $t$ is the number of days since launch.
The performance of our CTI correction algorithm on warm pixels is shown by the points in figure~\ref{fig:trap_density} with dashed error bars.
After correction with $n_\mathrm{iter}=6$ iterations, the residual trap density is well-fit by
$\rho_\mathrm{t}V_\mathrm{pix}=(0.032\pm0.023)-(0.24{\pm}5.46)$$\times$10$^{-5}(t-3374)$\,pixel$^{-1}$, where $t$ is the number of days since launch.
Comparing the constants in these two fits, we find that 98\% of the trail amplitude has been removed.

If we stop the iteration after $n_\mathrm{iter}=3$, the residual trap density (this is not shown in the figure) is fit by $\rho_\mathrm{t}V_\mathrm{pix}=(0.036\pm0.012)-(0.35{\pm}2.90)$$\times$10$^{-5}(t-3374)$\,pixel$^{-1}$.
This amplifies noise less (see section~\ref{sec:niter}), but produces almost as good a correction.
In some scientific contexts where statistical errors are sufficiently large to dwarf the CTI residual, we therefore advocate the use of lower $n_\mathrm{iter}=3$.

As an incidental part of our testing, we performed a sensitivity analysis by rerunning the CTI correction for six of the data sets from late 2012 with 1\% fewer charge traps.
This produced corrected images with a $(1.10\pm0.13)$\% smaller change in trail amplitude -- consistent with the expectation that both the amount of trailing and the amount of correction depends linearly on the density of traps. 

\begin{figure}
\begin{center}
\includegraphics[width=75mm]{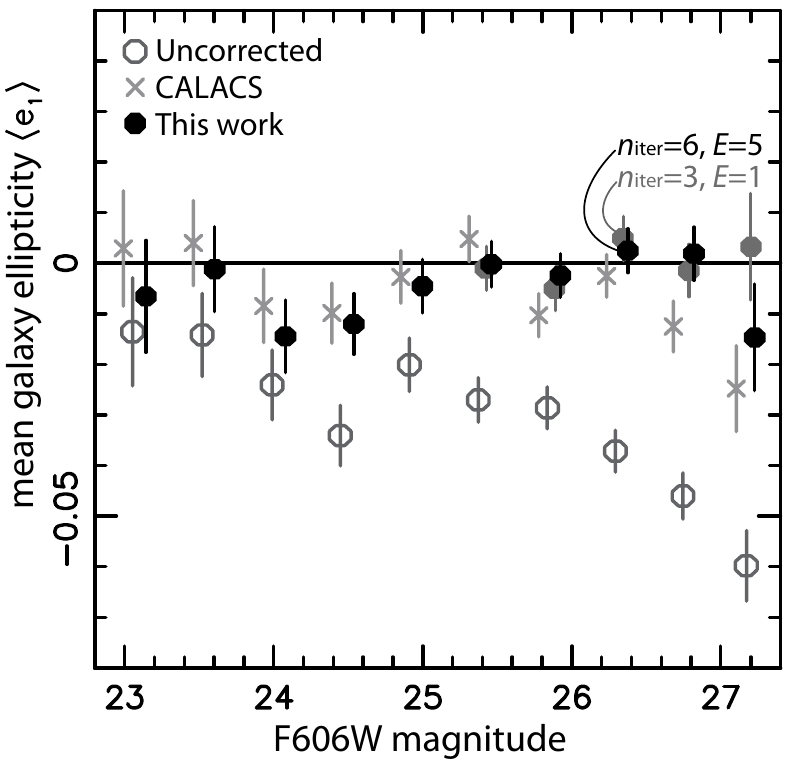}\vspace{-3mm}
\end{center}
\caption{
The measured shapes of faint galaxies in real {\sl HST} \acswfc\ imaging from 2011 and 2012, as a function of their brightness in the F606W band (different points use the same magnitude bins, but have been offset for clarity).
We show the mean ellipticity, averaged over a range of galaxy sizes and positions in the {\sl ACS} field of view.
Parallel CTI causes mean ellipticity $\langle e_1\rangle<0$ (net alignment along the $y$-axis).
Spurious elongation in the raw data is successfully removed when our CTI correction algorithm is applied as the first step of data analysis (the choice of $n_\mathrm{iter}$ does not affect bright galaxies but its effect on noise is just discernible for the faintest).
Note that the error bars are slightly underestimated because some galaxies are located in the overlapping regions of adjacent tiles, and contribute more than once.}
\label{fig:schrabbacksimple}
\end{figure}

\begin{figure*}
\begin{center}
\includegraphics[width=175mm]{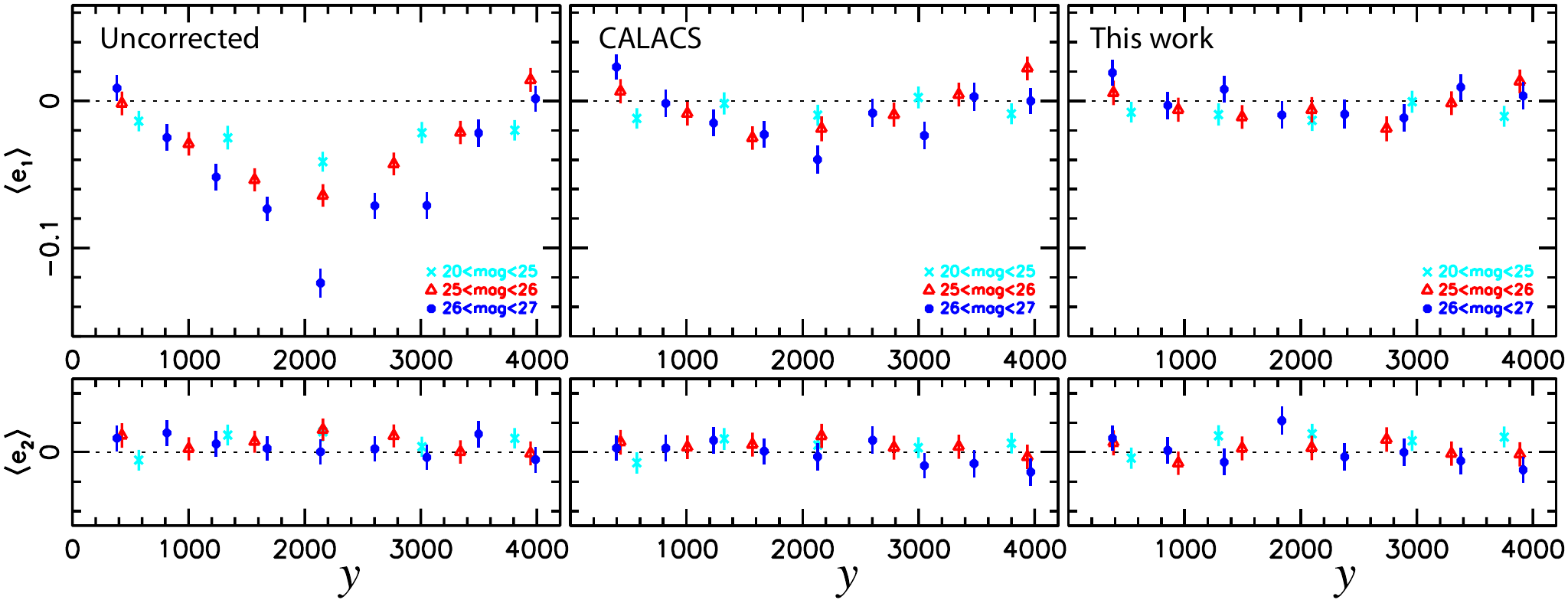}
\end{center}
\caption{
The measured shapes of faint galaxies in real {\sl HST} \acswfc\ imaging from 2011 and 2012, as a function of their $y$ position on the detector.
The left column shows raw data, and the other columns show analyses after correction with the public versions of different CTI correction packages (we use $n_\mathrm{iter}=3$, which limits noise amplification and slightly improves performance for the faintest galaxies).
The top (bottom) panels show the elongation of galaxies along (at $45^\circ$ to) the $y$ axis.
In this camera, readout nodes are at the bottom ($y=0$) and top (y=4096) of the detector, so the CTI is worst in the middle.
Note that the error bars are slightly underestimated because some galaxies are located in the overlapping regions of adjacent tiles, and contribute more than once.}
\label{fig:schrabbackthree}
\end{figure*}

\subsection{Extended sources}

Since the warm pixels used to check the performance of our algorithm above are the same pixels that were used to develop and calibrate the model, being able to correct them is necessary but not sufficient.
The performance of our CTI correction algorithm on independent images of real galaxies is demonstrated in figures~\ref{fig:schrabbacksimple} and \ref{fig:schrabbackthree}.
We also compare this to the performance of on-the-fly \citet{a10} processing, using CALACS PixelCTE 2012 v3.2 (with reference table {\sf w591643nj\_cte.fits}).
We used both packages to correct all the CANDELS F606W data taken in the COSMOS field between 2011 and 2012 \citep{candels1,candels2}, then measured galaxies' mean ellipticity $\langle e_1\rangle$ (see equation~\ref{eqn:rrge1}).
To correct for {\sl HST}'s PSF, we followed the method described in \cite{sch10}.
Positive values of ellipticity indicate that the galaxies' major axes tend to be aligned with the detector's $x$-axis, and negative values indicate alignment with the $y$-axis\footnote{For some science analyses, it may be possible to correct any residual at the catalogue level by simulating the data, the radiation damage, and the imperfect CTI removal \citep[\eg][]{cawley02,leauthaud07,chiaberge}. The correction will be smaller from the residual than from a raw measurement. However, accuracy relies on realism throughout the simulation, and every observable must be calibrated separately. It is therefore still preferable to meet requirements on CTI correction at the pixel level.}.
If there is no preferred direction in the Universe, the mean ellipticity ought to be consistent with zero.
This particular test is useful because it is self-contained (c.f.\ tests on galaxy positions or fluxes require comparison to `pristine' external imaging). 
It is also possible to perform this measurement very accurately, using software that has been developed to measure galaxy shapes at high precision for weak gravitational lensing analysis \citep[e.g.][]{hrev,mrev}.

Parallel CTI is dominant in the raw images, producing a spurious negative ellipticity of $\sim$0.12 for the faintest galaxies that are farthest from the readout register (averaging over all positions on the detector in figure~\ref{fig:schrabbacksimple} reduces this by half).
Our measurement accuracy will be limited by statistical noise owing to the finite number of galaxies in even this very large data set.
After correction, the measured ellipticities are consistent with zero, demonstrating that the CTI correction is successful.

The small residual variations from zero are probably due to imperfections in our model of the PSF. 
The PSF imprints itself onto the ellipticity of faint galaxies, and can only be removed to the accuracy of the PSF model. 
We are now actively re-reducing ACS images of dense stellar fields, to remove CTI trailing and improve our PSF model.

\section{Conclusions} \label{sec:conclusions}

The performance of the CCD detectors in {\sl HST} \acswfc\ has steadily degraded since they were placed in orbit in 2002.
They now exhibit  strong trailing due to Charge Transfer Inefficiency, including a spurious elongation of faint galaxies.
Such trailing now limits the application of {\sl HST} for many high precision analyses, including supernova photometry, gravitational lensing and proper motions.
If uncorrected, it may similarly limit science exploitation of future space missions including {\it Gaia} and {\it Euclid}.

\subsection{Accurate CTI correction for HST}

We have developed an algorithm to reproduce CTI trailing in {\sl HST} \acswfc\ (figure~\ref{fig:flowchart}).
This can also be used iteratively to remove CTI trailing (figure~\ref{fig:flowchart_inverse}).
Compared to the \cite{m10b} algorithm, which achieves $\sim$95\% correction on recent data in various scientific contexts, our new algorithm is 8 times faster and 2.5 times more accurate (successfully removing $\sim$98\% of the spurious trailing).
These advances are due to a combination of algorithmic improvements and recalibration upon in-orbit data up to early 2013.

In particular, it is now important that an $E>1$ CTI correction algorithm should be used for high precision applications.
In the words of \cite{a10}, the CTI trailing in {\sl ACS} is rapidly becoming `pathological', and more than a perturbation about the raw image.
An implementation of our code in C can be downloaded from \url{http://www.astro.uni-bonn.de/download/software/cte-tool/}.

\subsection{Meta-model from detailed literature survey}

As part of the process of optimising our CTI correction, we have compared (in detail) several successful algorithms from the literature that have been developed to model CCD readout.
Our main finding is the remarkable similarities between even completely independent codes.
Indeed, the codes can be summarised as subsets of an overarching model that takes several input parameters, described in table~\ref{tab:metamodel}.

In the trade-off between accuracy and speed of execution, the various codes each adopt a slightly different balance.
They already share several computer-science tricks to speed up runtime, and we have incorporated others. 
We have suggested a balance in these trades appropriate for {\sl HST} \acswfc.

In terms of physics, the codes' two main differences are: the procedure to determine which charge traps can (or are likely to) capture an electron, and the monitoring of which traps are full.
Theory \citep{srhsr,srhh} indicates that the probability for capture, $p_\mathrm{capture}(\mathbf{x},\sigma_t;n_e)$, should depend upon the traps' 3D position within a pixel $\mathbf{x}$ and capture cross-section $\sigma_t$.
The position can be reduced to a 1D variable, by carefully ordering the subpixel volume.
No working CTI correction algorithm has yet incorporated a full trapping model. 
\cite{seabroke} retain 3D structure, and the integrals over the pixel volume take too long to be practical;
\cite{m10b} and \cite{a10} collapse $p$ into a delta function; and 
\cite{short13} do not keep track of any dependence on $\mathbf{x}$.

\subsection{Possible directions for future work}

\subsubsection{Specific recommendations for {\em HST}} 

An algorithm with variable $E>1$ should be incorporated into the standard STScI data analysis pipeline.
To achieve 1\% CTI residuals, it will also be necessary to start correcting serial CTI.

The flexibility of \citet{a10}'s non-parametric form for $n_\mathrm{c}(n_e)$ bodes well for the translation of \acswfc\ code to {\it WFC3}/{\it UVIS} data (Anderson \etal\ in prep.), in which the sky background can be very low.
This has the ability to reproduce \citet{short13}'s $n_\mathrm{c}(n_e)$ at low $n_e$ (for a fixed clock speed and operating temperature), where experience from {\it Gaia} suggests a density-driven approach might be more suitable.
A density-driven model might also provide an elegant way to implement \cite{a10}'s observation that fast-release traps may be filled before slow-release traps, in the regime where $n_e<n_\mathrm{c}$.

\subsubsection{Recommendations for future missions} 

{\it Gaia} should consider using $n_\mathrm{levels}>1$ to monitor trap occupancy.
Even a modest increase to $n_\mathrm{levels}\sim50$ could improve accuracy, if the new trap levels are concentrated at low $n_e$ where they will have the most effect.
If the corresponding increase in run time is prohibitive, a similar saving could be achieved by accounting for only one species of trap instead of 4, but with a complex release profile to represent the mixture of species.

{\it Euclid} requires 99\% correction of CTI in the shapes of the faintest galaxies at the end of the mission \citep{m13,cropper13}, which is better than we have achieved for {\sl HST} in this paper.
\cite{euclid} had suggested noise whitening as a promising avenue to improve performance.
We find that the basic implementation of noise whitening does not have the desired effect.
However, steps have been taken to mitigate the problem via specifically optimised hardware, including slower readout to lower read noise (and avoid resonance between clock speed and trap release that affects observables of interest \citealt{rhodes10}) plus multi-level clocking \citep{triphase}.

Noise added by CTI correction can be further reduced if the 3D locations of individual charge traps in damaged detectors are known. 
An STScI program has been initiated to better measure these using long and short combinations of dark exposures taken during `internal' {\sl HST} time (\url{http://www.stsci.edu/hst/wfc3/tools/cte_tools}).
A potentially even greater benefit -- in both noise reduction and the calibration of software correction to unprecedented accuracy -- may be obtained from in-orbit trap pumping measurements \citep{janesick01,pumping}.
Where it is still possible to adapt readout electronics, we recommend that they should enhanced to enable in-orbit trap pumping.
Since CTI may otherwise seriously limit long-term science exploitation, modest investment in such hardware (and software to build models from trap pumping data) is likely to reap significant return.

We have not considered the effect of cosmic rays that hit during readout.
Their associated cloud of electrons will not be subject to as many pixel-to-pixel transfers as others, so will not be as trailed.
Identifying these cosmic rays will be especially important for missions with continuous or slow readout.
\cite{a10} suggest an iterative procedure to identify and avoid under-correcting them.

\subsubsection{Recommendations for future software} 

We have explored most of the physics choices and the tradeoffs between speed and accuracy that are summarised in table~\ref{tab:metamodel}.
However, a starting point for future software development (especially for new detectors) should probably be a meta-code that can easily take any or all of the options from table~\ref{tab:metamodel} as input parameters.
Code optimisation starting from that point may further improve accuracy, over a wider range of operating conditions.

A useful addition to readout algorithms would be a (more sophisticated) treatment of multi-phase clocking.
Multi-level clocking \citep{triphase} could then be modelled by adjusting the direction travelled by released electrons.
Multiple phases could probably be implemented in $n_\mathrm{phases}$ static events, but the dwell times should be allowed to differ between phases, and the changing electrostatic potentials may need different full well depth and filling parameters.
If the charge clouds physically overlap in adjacent phases, it may even be necessary to use full 3D modelling like \citet{hall10} -- or even allow charge to flow through the silicon at different speeds and densities at different positions.
However, this would become very slow.

The most important improvement for iterative CTI correction will be a method to circumvent the limitations of read noise.
Read noise is added after charge transfer, so is never trailed.
Algorithms to undo the trailing may perfectly correct the underlying image, but `over-correct' the read noise.
\cite{a10} proposed a tuneable high-/low-pass filter to identify some component of the read noise; we proposed a scheme to add correlated noise in such a way that the total noise becomes white.
Our method works in other contexts \citep{huff11} but did not have the desired effect here.
Estimators for galaxy shapes, sizes, etc.\ that are carefully constructed to average out background noise, are unaffected by adding more noise to a final image (by construction). 
However, even such carefully-obtained measurements yield the wrong values if read noise is added between the addition and the removal of CTI trailing.

We remain unsure as to the exact mechanism through which read noise prevents perfect measurements.
{\it Something} fundamental about the image is being irrevocably lost by the over-correction of read noise --- but it is not merely `amplification' of read noise \citep{a10} or even `anti-correlation' of read noise (upon which assumption was our method based).
We speculate that it may be due to a more subtle asymmetry in properties of the noise after correction, even though the read noise was Gaussian when it was added.
Because ${\rm d}n_\mathrm{c}/{\rm d}n_e<1$, the number of traps to which each marginal electron gets exposed is smaller than the last. 
We therefore note that
\begin{itemize}
\item
Positive fluctuations in read noise peaks are trailed (and untrailed) slightly less than negative fluctuations.
\item
Read noise (or photon shot noise) on top of an object is untrailed less than those in the wings of an object or an area of blank sky. 
\end{itemize}
There are also two asymmetries from one side of an object to the other
\begin{itemize}
\item
Read noise on the leading side of a source is untrailed more than read noise on the (shadowed) lee side.
\item
Read noise on the side of an object closest to the readout amplifier is (very slightly) untrailed less than read noise far from the amplifier.
\end{itemize}
Determining which of these (or other) effects drives the current limitation is now beyond the scope of this paper.
However, we recommend that it will be worth investigating the asymmetries of post-correction read noise in future work.

\section*{Acknowledgments}

The authors thank Alex Short, Jay Anderson and Norman Grogin for sharing and discussing their CCD readout algorithms, Tom Kitching for help with noise whitening code, and Andrew Clarke for providing the data for figure~\ref{fig:silvaco}.
We also thank Roger Smith, Jason Gow, Neil Murray and the rest of the Euclid CCD working group for more conversations about CTI. 

RM is supported by a Royal Society University Research Fellowship, European Research Council grant MIRG-CT-208994, and Philip Leverhulme Prize PLP-2011-003. 
TS acknowledges support from the German Federal Ministry of Economics and Technology (BMWi) provided via DLR under project 50OR1210.
OC and OM are supported by BMWi under DLR-Grant 50QE1103.
HI is supported by STFC grant ST/K003305/1.

This work was supported in part by the National Science Foundation under Grant No. PHYS-1066293 and the hospitality of the Aspen Center for Physics.
This work used the DiRAC Data Centric system at Durham University, operated by the Institute for Computational Cosmology on behalf of the STFC DiRAC HPC Facility (www.dirac.ac.uk). This equipment was funded by BIS National E-infrastructure capital grant ST/K00042X/1, STFC capital grant ST/H008519/1, and STFC DiRAC Operations grant ST/K003267/1 and Durham University. DiRAC is part of the National E-infrastructure. 

\noindent {\it Facilities:}
This paper uses data from observations with the NASA/ESA {\em Hubble Space Telescope}, obtained at the Space Telescope
Science Institute, which is operated by AURA Inc, under NASA contract NAS 5-26555.
Data were used from the following programmes (P.I.\ names listed in brackets), and are available in the online {\sl HST} archive:
GO-9075 (S.~Perlmutter), 
GO-9822 (N.~Scoville), 
GO-10092 (N.~Scoville), 
GO-10268 (D.~Trilling), 
GO-10496 (S.~Perlmutter), 
GO-10572 (T.~Lauer), 
GO-10896 (P.~Kalas), 
GO-10917 (D.~Fox), 
GO-11563 (G.~Illingworth),
GO-11600 (B.~Weiner),
GO-11663 (M.~Brodwin),
GO-11877 (M.~Lallo),
GO-12246 (C.~Stubbs),
GO-12444 (S.~Faber), 
GO-12477 (F.~High) and
GO-12994 (A.~Gonzalez).

\bsp
\label{lastpage}

\end{document}